\documentclass[acmtog]{acmart}

\usepackage{booktabs} 

\usepackage[ruled]{algorithm2e} 

\SetAlFnt{\small}
\SetAlCapFnt{\small}
\SetAlCapNameFnt{\small}
\SetAlCapHSkip{0pt}

\acmJournal{TOG}
\setcopyright{acmlicensed}\acmJournal{TOG}
\acmYear{2020}\acmVolume{39}\acmNumber{6}\acmArticle{265}\acmMonth{12} \acmDOI{10.1145/3414685.3417789}

\usepackage{wrapfig}
\usepackage{nicefrac}
\usepackage{mathtools}
\usepackage{graphicx}
\usepackage{booktabs} 
\usepackage{combelow} 
\usepackage{tabularx} 
\usepackage{colortbl} 

\usepackage{manfnt} 

\usepackage{amsmath}
\usepackage{amsfonts}
\usepackage{amssymb}
\usepackage{amsbsy}

\usepackage[utf8x]{inputenc}

\definecolor{derekBlue}{RGB}{144,210,236}
\definecolor{honglinPurple}{RGB}{115, 113, 235}
\definecolor{derekTableBlue}{RGB}{189,235,252}
\definecolor{reviseBlue}{RGB}{65, 105, 225}
\definecolor{iglGreen}{RGB}{153,203,67}
\definecolor{coralRed}{RGB}{250,114,104}
\definecolor{gray}{RGB}{180,180,180}
\definecolor{alecGreen}{RGB}{51,204,51}

\SetCommentSty{commentFont}
\SetNlSty{textbf}{}{.}

\newcommand{\revise}[1]{{#1}}

\newcommand{\refequ}[1] {Eq.~\ref{equ:#1}}

\newcommand{\reffig}[1] {Fig.~\ref{fig:#1}}

\newcommand{\refsec}[1] {Sec.~\ref{sec:#1}}
\newcommand{\refapp}[1] {App.~\ref{app:#1}}

\newcommand{\refthm}[1] {Theorem~\ref{thm:#1}}

\DeclareMathOperator*{\argmin}{arg\,min}
\DeclareMathOperator*{\minimize}{minimize}
\newtheorem{theorem}{Theorem}

\newcommand{\R}{\mathbb{R}} 
\renewcommand{\C}{\mathcal{C}} 
\newcommand{\E}{\mathcal{E}} 

\newcommand{\X}{\mX} 
\newcommand{\x}{\vx} 
\newcommand{\xE}{\vx_{\E}} 

\newcommand{\Z}{\mZ} 
\newcommand{\rZ}{\widetilde{\mZ}} 
\newcommand{\z}{\vz} 
\newcommand{\rz}{\tilde{\vz}} 
\newcommand{\y}{\vy} 

\renewcommand{\S}{\mathbb{S}} 

\renewcommand{\i}{i} 

\renewcommand{\P}{\mP} 
\newcommand{\invP}{\mP^{\text{-}1}} 
\newcommand{\Q}{\mQ} 
\newcommand{\K}{\mK} 
\newcommand{\rK}{\widetilde{\mK}} 
\newcommand{\s}{s} 

\newcommand{\ie}{\textit{i.e.}}
\newcommand{\eg}{\textit{e.g.}}

\newcommand{\vecFont}[1]{\mathbf{#1}}

\def\vc{{\vecFont{c}}}

\def\ve{{\vecFont{e}}}

\def\vr{{\vecFont{r}}}
\def\vs{{\vecFont{s}}}

\def\vu{{\vecFont{u}}}
\def\vv{{\vecFont{v}}}
\def\vw{{\vecFont{w}}}
\def\vx{{\vecFont{x}}}
\def\vy{{\vecFont{y}}}
\def\vz{{\vecFont{z}}}

\newcommand{\matFont}[1]{\mathbf{#1}}
\def\mA{{\matFont{A}}}
\def\mB{{\matFont{B}}}
\def\mC{{\matFont{C}}}
\def\mD{{\matFont{D}}}
\def\mE{{\matFont{E}}}

\def\mG{{\matFont{G}}}

\def\mI{{\matFont{I}}}

\def\mK{{\matFont{K}}}
\def\mL{{\matFont{L}}}
\def\mM{{\matFont{M}}}

\def\mP{{\matFont{P}}}
\def\mQ{{\matFont{Q}}}
\def\mR{{\matFont{R}}}

\def\mU{{\matFont{U}}}
\def\mV{{\matFont{V}}}
\def\mW{{\matFont{W}}}
\def\mX{{\matFont{X}}}
\def\mY{{\matFont{Y}}}
\def\mZ{{\matFont{Z}}}

\begin{document}

\title{Chordal Decomposition for Spectral Coarsening}


\author{Honglin Chen}
\affiliation{%
  \institution{University of Toronto}
  \streetaddress{40 St. George Street}
  \city{Toronto}
  \state{ON}
  \country{Canada}
  \postcode{M5S 2E4}
}
\email{chl9797@cs.toronto.edu}

\author{Hsueh-Ti Derek Liu}
\affiliation{%
  \institution{University of Toronto}
  \streetaddress{40 St. George Street}
  \city{Toronto}
  \state{ON}
  \country{Canada}
  \postcode{M5S 2E4}
}
\email{hsuehtil@cs.toronto.edu}

\author{Alec Jacobson}
\affiliation{%
 \institution{University of Toronto}
 \streetaddress{40 St. George Street}
 \city{Toronto}
 \state{ON}
 \country{Canada}
 \postcode{M5S 2E4}
}
\email{jacobson@cs.toronto.edu}

\author{David I.W. Levin}
\affiliation{%
 \institution{University of Toronto}
 \streetaddress{40 St. George Street}
 \city{Toronto}
 \state{ON}
 \country{Canada}
 \postcode{M5S 2E4}
}
\email{diwlevin@cs.toronto.edu}



\begin{teaserfigure}
    \includegraphics[width=\linewidth]{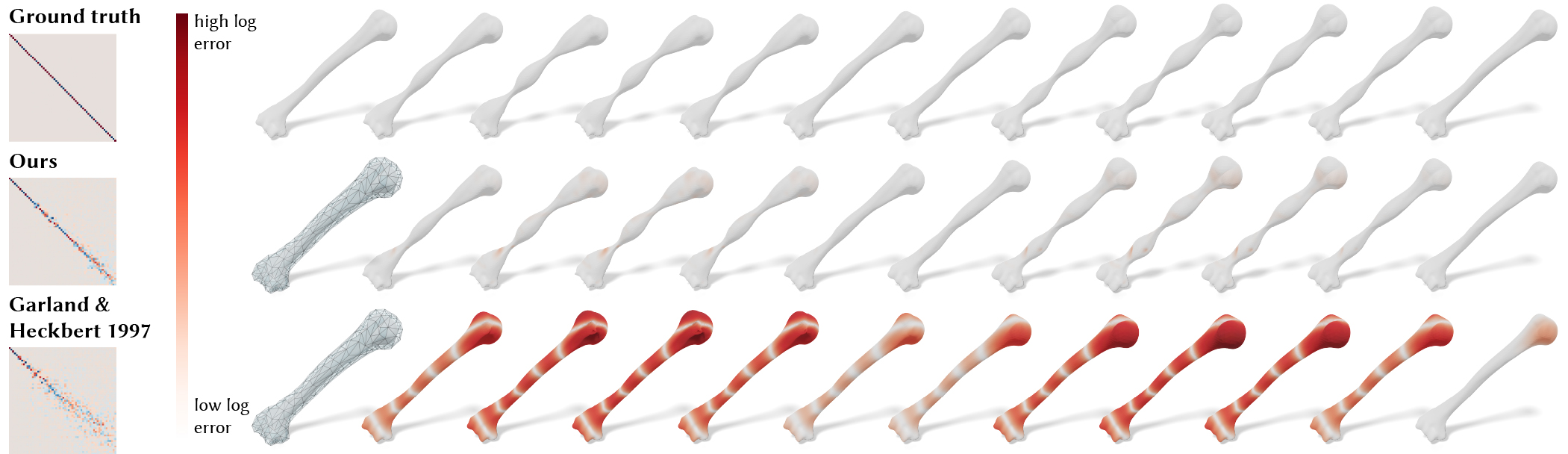}
    \caption{We approximate the vibration modes of the cotangent Laplacian derived from the ground truth high-resolution mesh (top) using a coarse mesh with 250 vertices (the transparent cages on the left). A classical decimation method \cite{GarlandH97} (bottom) preserves the appearance but fails in preserving the ground truth vibration modes. Our chordal spectral coarsening detaches the differential operator from the mesh, enabling one to optimize the operator independently to preserve the vibration modes (middle), without altering the coarse vertices. By visualizing the inner product matrices between vibration modes on the left, we show our approach leads to a result closer to the ground truth. Here we visualize the 9-th vibration mode with its frequency.}
    \label{fig:teaser}
\end{teaserfigure}

\begin{abstract}
    We introduce a novel solver to significantly reduce the size of a geometric operator while preserving its spectral properties at the lowest frequencies.
    We use chordal decomposition to formulate a convex optimization problem which allows the user to control the operator sparsity pattern.
    This allows for a trade-off between the spectral accuracy of the operator and the cost of its application.
    We efficiently minimize the energy with a change of variables and achieve \revise{state-of-the-art} results on spectral coarsening.
    Our solver further enables novel applications including volume-to-surface approximation and detaching the operator from the mesh, \revise{\ie,} one can produce a mesh tailor-made for visualization and optimize an operator separately for computation.
\end{abstract}
\maketitle




\vspace*{-2pt}
\section{Introduction}
Discrete operators, such as the cotangent Laplacian, the Hessian of mesh energies, and the stiffness matrix in physics-based simulations, are ubiquitous in geometry processing. 
Many of these operators are represented by sparse positive semi-definite (PSD) matrices. 
These matrices are often constructed by looping over the elements of a discretized domain. 
When defined on a high-resolution domain, those matrices are computationally expensive to use, even if the final result only requires low frequency information. 

Recent methods show that it is possible to simplify a discrete operator while preserving its spectral properties and matrix characteristics, such as positive semi-definiteness, avoiding the pitfalls of the na\"ive ``decimate and reconstruct'' approach. 
However, previous methods required the solution of a non-convex optimization problem, the solution to which sacrificed matrix sparsity. 

In this paper, we overcome these challenges by applying the \emph{chordal decomposition}.
In contrast to the previous non-convex formulation, our method is now convex and can freely control the output sparsity, outperforming existing approaches for spectral coarsening and simplification. 
Our approach further enables novel applications on optimizing the operator independently to preserve some desired properties for computation without changing the mesh vertices. 
In \reffig{teaser}, we first decimate the model and optimize the operator independently to preserve the spectral properties of the cotangent Laplacian.  Our approach achieves a higher quality approximation of the vibration modes of the high-resolution mesh compared to previous approaches.

%
%
%
\begin{wrapfigure}[11]{r}{0.4\linewidth}
    \vspace*{-1\intextsep}
    \hspace*{-0.5\columnsep}
    \begin{minipage}[b]{\linewidth}
    \includegraphics[width=\linewidth, trim={0mm 4mm 0mm 0mm}]{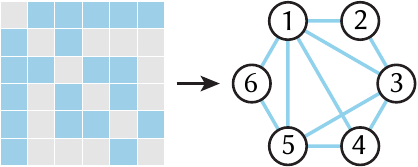}
    \caption{Given a sparse matrix \revise{(left)} where blue denotes non-zeros and gray denotes zeros, we can view the sparsity pattern as a graph (right) and then apply theorems of chordal graphs. }
    \label{fig:chordalGraph}
    \end{minipage}
\end{wrapfigure}
\begin{figure}
    \centering
    \includegraphics[width=3.33in]{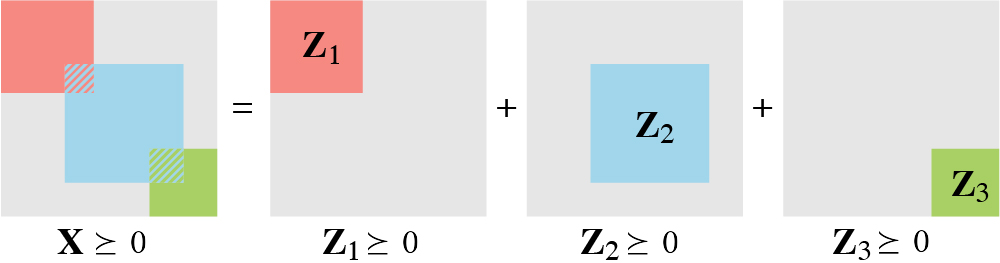}
    \caption{We use chordal decomposition to split a large sparse PSD constraint on $\X$ (left) into multiple small dense PSD constraints on $\Z_\i$ (right), where we use $\cdot \succeq 0$ to denote the PSD constraint. This enables us to be more efficient in handling optimization problems that involve sparse PSD matrix constraints. }
    \label{fig:multiPSD}
    \vspace{-12pt}
\end{figure}

By viewing the sparsity pattern of a matrix as a graph (see \reffig{chordalGraph}), one can utilize theories of chordal graphs to decompose a \emph{sparse} matrix into a set of small \emph{dense} matrices. 
This decomposition enables one to satisfy the sparse PSD constraint by projecting each small dense matrix to PSD in parallel (see \reffig{multiPSD}). Such techniques have long been applied in the creation of efficient solvers for Semidefinite Programming (SDP). 
Here we generalize these notions to the spectral coarsening problem which leads to an accelerated solver that is faster, more accurate \revise{and with better sparsity control} than \revise{the} previous state-of-the-art. 
\revise{Our main contribution is an algorithm for projecting general sparse matrices to PSD ones using chordal decomposition in the context of spectral coarsening. }
\vspace*{-6pt}
\section{Related Work}
Spectral preservation is a widely studied topic in optimization and numerical methods. Below we outline the most salient related works from these areas, as well as recent developments in computer graphics and geometry processing. 

\subsection{Chordal Graphs in Sparse Matrix Optimization}
Chordal graphs have been playing an important role in sparse matrix computation for decades \cite{blair1993introduction, vandenberghe2015chordal}. 
\citet{FukudaKMN01} and \citet{NakataFF03} introduce a generic framework to accelerate interior-point methods for solving large sparse SDPs. Their key idea is to exploit the sparsity of the matrix and the properties of chordal graphs \cite{grone1984positive} to decompose a large sparse matrix variable into multiple small dense ones. In the later literature, this is often called the \emph{chordal decomposition}. 
Since then, this framework has been greatly improved by \cite{Burer03, Srijuntongsiri2004, AndersenDV10, sparseCOLO, Sun2014Decomposition}. 
The idea of chordal decomposition has also been incorporated with other optimization methods.
For instance, \citet{sun2015decomposition} combine chordal decomposition with projected gradient and the Douglas–Rachford algorithms for sparse matrix \revise{nearness} and completion problems. 
\citet{Zhen2017ADMMSDP, Zhen2020ADMMSDP} incorporate this idea to the alternating direction method of multipliers (ADMM) for solving SDPs.
These chordal-based solvers have also been deployed to nonlinear matrix inequalities \cite{KimKMY11}, the optimal power flow \cite{MadaniKL15}, controller synthesis \cite{ZhengMP18} and sum-of-squares problems \cite{zheng2017exploiting, ZhengFP19}.

Recently, \citet{Maron2016} formulate the point cloud registration problem into a SDP and use chordal decomposition to accelerate the computation.
However, their method only supports matrices with a chordal sparsity pattern already, which is not applicable to our problem because most discrete operators are not chordal.
In contrast, we utilize the ideas from \cite{sun2015decomposition} to handle any sparsity pattern of choice, and the strategies in \cite{Zhen2017ADMMSDP,Zhen2020ADMMSDP} to develop a chordal ADMM solver for the spectral coarsening energy \cite{liu2019spectral}. 
We exploit the fact that many discrete operators are sparse and symmetric to perform a change of variables to significantly reduce the computational cost.
We demonstrate that chordal decomposition is not only suitable for large scale SDPs, but also for problems in graphics that involve sparse PSD matrix variables.


\subsection{Geometry Coarsening} 
Geometric coarsening has been extensively studied in computer graphics with the aims of preserving different geometric and \revise{physical} properties. One class of methods focuses on preserving the appearance of a mesh for rendering purposes. Some prominent early examples include mesh optimization \cite{HoppeDDMS93, CohenSteinerAD04}, mesh decimation \cite{GarlandH97}, progressive refinement \cite{Hoppe96, Hoppe97}, and approaches based on parameterization \cite{CohenMO03}. We refer readers to \cite{CignoniMS98} for an overview and comparison of appearance-preserving simplification.
Beyond preserving the appearance, these techniques have also been extended to preserve the texture information of a shape \cite{GarlandH98, LuLWGHCHL20}.
\citet{LiFZ15} add modal displacement as part of \revise{the} decimation metric to better preserve the acoustic transfer of a shape. 

\paragraph{Numerical coarsening in simulation} Coarsening the geometry may alter the material properties and lead to \emph{numerical stiffening} in simulations. 
\citet{KharevychMOD09} propose a method to adjust the elasticity tensor of each element on a coarse mesh to approximate the dynamics of the original high-resolution mesh. 
In a similar spirit, \citet{Matusik15} use a data-driven lookup approach to reduce the error incurred by coarsening. 
To better capture vibration, \citet{ChenLMK17} address the numerical stiffening by simply rescaling the Young's modulus of the coarse model to match the lowest frequencies to its high-resolution counterpart.
\citet{ChenLKAP19} extend this idea to re-fit the eigenvalues iteratively at each time step.
\citet{ChenJH2018} propose to construct matrix-valued and discontinuous basis functions by solving a large amount of local quadratic constrained optimizations.
Other recent approaches have included the wavelet approaches.
\citet{Owhadi17} introduces a hierarchical construction of operator-adapted basis \revise{functions} and their associated wavelets for scalar-valued PDE.
The operator-adapted wavelets have been extended to differential forms \cite{BudninskiyOD19} and to vector-valued equations \cite{Chen2019bases} which is then applied to fast simulation of heterogeneous materials with locally supported basis functions.
Different from \cite{ChenJH2018} and \cite{Chen2019bases} which increase the degrees of freedom (DOF) by using matrix-valued shape functions, our method can support more DOF by directly controlling the sparsity pattern of the scalar-valued matrix. 
Moreover, our method can also preserve the spectral properties using the same DOF and sparsity pattern.

\revise{\paragraph{Spectral graph coarsening in machine learning} Spectral-preserving graph reduction has been an active field in machine learning. 
\citet{DBLP:journals/corr/Spectral2018} introduce a scalable spectral graph reduction method for scalable graph partitioning and data visualization based on node aggregation and graph sparsification.
\citet{DBLP:conf/aistats/JinLJ20} propose two methods for spectral graph coarsening based on iterative merging and k-means clustering, respectively. 
Various other approaches have also been recently adopted to coarsen a graph in a spectral-preserving way, including randomized edge contraction \cite{DBLP:conf/icml/LoukasV18}, local variation algorithm \cite{DBLP:journals/jmlr/Loukas19} and probabilistic algorithm \cite{NIPS2019_Spectrum}. 
In contrast to these combinatorial methods which focus more on optimizing the sparsity pattern, our algebraic approach enables one to further optimize over a specific sparsity pattern based on a convex formulation. }

\paragraph{Spectral coarsening in geometry processing} Recently several approaches consider coarsening a geometry while preserving its spectral properties, namely eigenvalues and eigenvectors of the operators. 
\citet{ztireliAG10} compute samples on a manifold surface in order to preserve the spectrum of the Laplace operator.
\citet{NasikunBH18} use a combination of Poisson disk sampling and local polynomial bases to efficiently solve an approximate Laplacian eigenvalue problem of a mesh.
Beyond the Laplace operator, \citet{liu2019spectral} propose an algebraic approach to coarsen common geometric operators while preserving spectral properties.
\citet{lescoat2020spectral} extend the formulation to achieve spectral-preserving mesh simplification.
Our approach is purely algebraic. Our convex formulation leads us to have better spectral preservation compared to the similar algebraic approach \cite{liu2019spectral} in spectral coarsening.
Our flexibility in controlling the sparsity allows us to post-process the results of spectral simplification \cite{lescoat2020spectral} and further improve its quality. 
In addition, we enable a novel application which independently optimizes the operator for computation purposes and the mesh vertices for preserving the appearance (see \reffig{teaser}).

\section{Background}

The description of our method depends on manipulating variables that represent sparse matrices. 
%
%
Throughout the paper, we use $\P$ to denote \emph{selection matrices}, and use subscripts to differentiate between them. In practice, given a subset $\s$, $\P_\s$ is a sparse matrix defined as
\begin{align}\label{equ:selectionMatrix}
	(\P_\s)_{jk} = 
	\begin{cases}
		1, & k = \s(j),\\
		0, & \text{otherwise}.
	\end{cases}
\end{align}
Let $\x$ be a vector and $\z = \x(\s)$ be a sub-vector of $\x$. Selecting a subset from $\x$ can be achieved by a sparse matrix multiplication $\z = \P_\s \x$. Mapping the elements from $\z$ to a bigger vector $\vy$ can be achieved with $\vy = \P_\s^\top \z$
\begin{align*}
	\underbrace{\begin{bmatrix}
		c_1 \\ c_2 \\ c_4
	\end{bmatrix}}_\z
	= 
	\underbrace{\begin{bmatrix}
		1 & 0 & 0 & 0 \\
		0 & 1 & 0 & 0 \\
		0 & 0 & 0 & 1 
	\end{bmatrix}}_{\P_\s}
	\underbrace{\begin{bmatrix}
		c_1 \\ c_2 \\ c_3 \\ c_4
	\end{bmatrix}}_\x, \quad 
	\underbrace{\begin{bmatrix}
		c_1 \\ c_2 \\ 0 \\ c_4 
	\end{bmatrix}}_\vy
	= 
	\underbrace{\begin{bmatrix}
		1 & 0 & 0  \\
		0 & 1 & 0 \\
		0 & 0 & 0 \\
		0 & 0 & 1 
	\end{bmatrix}}_{\P^\top_\s}
	\underbrace{\begin{bmatrix}
		c_1 \\ c_2 \\ c_4 
	\end{bmatrix}}_\z.
\end{align*}

Let $\X$ be an $n$-by-$n$ matrix, we use $\s$ to denote a subset of row and column indices into $\X$. $\Z = \X(\s,\s)$ creates a matrix $\Z$ that is of size $|\s|$-by-$|\s|$ and contains all values in $\X(\s,\s)$. 
We can compactly describe this operation using selection matrix $\P_\s$ as $\Z = \P_\s \X \P_\s^\top$
\begin{align*}
	\underbrace{\begin{bmatrix}
		c_{11} & c_{13} \\ c_{31} & c_{33}
	\end{bmatrix}}_\Z
	= 
	\underbrace{\begin{bmatrix}
		1 & 0 & 0 \\
		0 & 0 & 1 \\
	\end{bmatrix}}_{\P_\s}
	\underbrace{\begin{bmatrix}
		c_{11} & c_{12} & c_{13} \\ c_{21} & c_{22} & c_{23} \\ c_{31} & c_{32} & c_{33}
	\end{bmatrix}}_\X
	\underbrace{\begin{bmatrix}
		1 & 0  \\
		0 & 0 \\
		0 & 1
	\end{bmatrix}}_{\P_\s^\top}.
\end{align*}
Similarly, we can map elements in $\Z$ back to $\mY$ via $\mY = \P_\s^\top \Z \P_\s$
\begin{align*}
	\underbrace{\begin{bmatrix}
		c_{11} & 0 & c_{13} \\ 0 & 0 & 0 \\ c_{31} & 0 & c_{33}
	\end{bmatrix}}_\mY
	= 
	\underbrace{\begin{bmatrix}
		1 & 0  \\
		0 & 0 \\
		0 & 1
	\end{bmatrix}}_{\P_\s^\top}
	\underbrace{\begin{bmatrix}
		c_{11} & c_{13} \\ c_{31} & c_{33}
	\end{bmatrix}}_\Z
	\underbrace{\begin{bmatrix}
		1 & 0 & 0 \\
		0 & 0 & 1 \\
	\end{bmatrix}}_{\P_\s}.
\end{align*}

\subsection{Chordal Decomposition}\label{sec:chordalDecomposition}
A \emph{chordal graph} is an undirected graph in which for every cycle of length greater than three, there is an edge between nonconsecutive vertices in the cycle. 
Chordal graphs have drawn attention since the 1950s because a handful of NP-complete graph problems can be solved in polynomial time if the graph is chordal. 
Chordal graphs also received interests from the optimization community for solving sparse SDPs, combinatorial optimization, and Cholesky factorization. 
We refer readers to \cite{vandenberghe2015chordal} for a survey of chordal graphs in optimization. We focus on its application to problems that involve sparse PSD matrices constraints, specifically arising from geometry processing.

An $n$-by-$n$ symmetric matrix $\X$ has \emph{chordal sparsity} pattern $\C \in \{0,1\}^{n \times n}$ if the graph induced by $\C$ is a chordal graph. 
\begin{figure}
	\centering
	\includegraphics[width=3.33in]{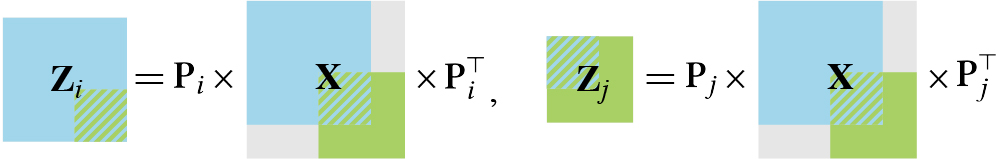}
	\caption{Chordal decomposition decomposes the matrix $\X$ into a set of maximal clique matrices $\Z_\i$. We can extract each clique matrix via $\Z_\i = \P_\i \X \P_\i^\top$.}
	\label{fig:chordalDecomposition}
	\vspace{-5pt}
\end{figure}
The key theorem that supports our method is
%
\begin{theorem} \label{thm:chordalPSD}
	(\cite{agler1988positive, kakimura2010direct}) Let $\X$ be a $n$-by-$n$ symmetric matrix with chordal sparsity, and let $\{ \Z_1, \Z_2, \cdots, \Z_p \}$ be a set of its $p$ \emph{clique matrices}. Then $\X$ is PSD if and only if it can be expressed as
	\begin{align}\label{equ:chordalDecomposition}
		\X = \sum_{\i=1}^p \P_\i^\top \Z_\i \P_\i
	\end{align}
	with all $\Z_\i$ being PSD.
\end{theorem}
\noindent
where a \emph{clique} is a subset of vertices such that every two distinct vertices in the clique are adjacent to each other, thus a \emph{clique matrix} is a dense matrix of the size of a clique. We use $\Z_\i$ to represent the $\i$th clique matrix, $\P_\i$ as the selection matrix to the $\i$th clique set. This decomposition from $\X$ to a set of clique matrices is called the \emph{chordal decomposition} (see \reffig{chordalDecomposition}), which has been applied to many recent SDP solvers.

%
\paragraph*{Vectorization} In practice, the ``sandwich'' format $\P_\i^\top \Z_\i \P_\i$ is not always easy to work with. It is often more desirable to \emph{vectorize} a matrix by concatenating the columns of the matrix into a vector (see the inset). We can re-write the vectorized chordal decomposition as 
\begin{align}\label{equ:vectorizedChordalDecomposition}
	\text{vec}(\X) = \sum_{\i=1}^p \text{vec}(\P_\i^\top \Z_i \P_\i) =  \sum_{\i=1}^p \underbrace{(\P_\i^\top \otimes \P_\i^\top)}_{\K_\i}\ \text{vec}(\Z_\i),
\end{align}
\begin{wrapfigure}[7]{r}{1.33in}
	\raggedleft
    \vspace{-7pt}
	\hspace*{-0.7\columnsep}
	\includegraphics[width=1.31in, trim={6mm 0mm 1mm 0mm}]{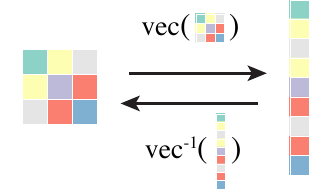} 
	\label{fig:vectorization}
\end{wrapfigure} 
where we use $\text{vec}(\cdot)$ to denote the vectorization, with its inverse $\text{vec}^{\text{-}1}$ (see the inset), and $\otimes$ to denote the Kronecker product. Intuitively, $\K_\i$ acts like the transpose of a selection matrix, putting elements in $\text{vec}(\Z_\i)$ back to $\text{vec}(\X)$. 

\subsection{Chordal Extension} \label{sec:chordalExtension}
In practice, a majority of matrices we encounter in geometry processing do not naturally have chordal sparsity patterns, which makes \refthm{chordalPSD} inapplicable. 
In response, we follow the idea in \cite{sun2015decomposition} to first perform a \emph{chordal extension} to transform the original non-chordal sparsity $\E$ to a chordal sparsity pattern $\C$ (see the inset). 
We maintain $\E$ by adding equality constraints to enforce \revise{new fill-in elements arising from the extension} to be zeros
\begin{align}\label{equ:chordalExtension}
	\X \in \S^n_\E\ \Rightarrow \ 
	\begin{aligned}
		&\X \in \S^n_\C,\\
		&\X_{jk} = 0, \quad \forall (j,k) \in \C \backslash \E,
	\end{aligned}
\end{align}
\begin{wrapfigure}[5]{r}{1.33in}
	\raggedleft
    \vspace{-10pt}
	\hspace*{-0.7\columnsep}
	\includegraphics[width=1.31in, trim={6mm 0mm 1mm 0mm}]{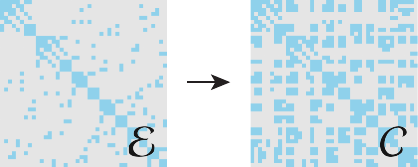} 
	\label{fig:chordalExtension}
\end{wrapfigure} 
where $\S^n_\E$ and $\S^n_\C$ denote $n$-by-$n$ symmetric matrices with sparsity patterns $\E$ and $\C$, respectively. We use $\C \backslash \E$ to denote the entries that exist in $\C$, but not in $\E$. 
\revise{Chordal extension adds degrees of freedom to our optimization problem. Our zero constraints enforce that, at a particular new fill-in entry, the sum of projected dense matrices must equal zero, not that each dense matrix must contribute a zero value to that entry. }
Computing the \emph{minimum} chordal extension, where the number of fill-in edges is minimized, is NP-complete \cite{Yannakakis1981}. However, finding a \emph{minimal} chordal extension can be solved in polynomial time \cite{Heggernes06}. 

Notice that \refthm{chordalPSD} also has a dual format $\mY_\i = \P_\i \X \P_\i^\top$.
If $\X$ has the chordal sparsity, this dual formulation can guarantee $\X$ to be PSD by ensuring all $\mY_\i$ being PSD, proved by the Theorem 7 in \cite{grone1984positive}. 
However, this dual formulation cannot guarantee $\X$ to be PSD if the matrix $\X$ does not have chordal sparsity (see Sec. 3.2.2 in \cite{sun2015thesis}). Thus we build our algorithm surrounding \refthm{chordalPSD}.

\section{Method} \label{sec:method}
\begin{figure}
    \centering
    \includegraphics[width=3.33in]{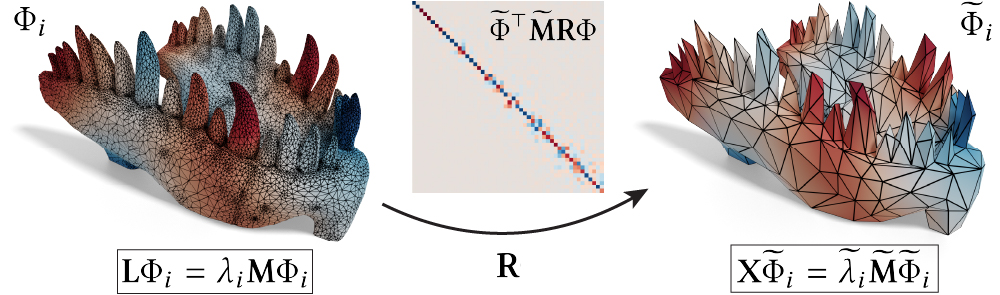}
    \caption{We visualize the spectral preservation using the inner product matrix (middle) between the restricted eigenvectors $\mR\Phi$ of the original operator $\mL$ to the coarsened domain and the eigenvectors $\widetilde{\Phi}$ of the coarsened operator $\mX$. Due to the orthonormality, the ground truth should be a diagonal matrix of 1 and -1 \revise{(denoted by red and blue, respectively)}. The closer the matrix to a diagonal matrix, the better the preservation of eigenvectors. We use $\mM$ and $\widetilde{\mM}$ to denote the mass matrices of the original and the coarsened meshes respectively. }
    \label{fig:innerProductMat}
    \vspace{-5pt}
\end{figure}

The goal of spectral coarsening is to reduce the size of a discrete operator, derived from a 3D shape, while preserving its spectral properties. \citet{liu2019spectral} show that it is possible to have a significant reduction without affecting the low-frequency eigenvectors and eigenvalues. 
They visualize the preservation of spectral properties with the inner product matrix between eigenvectors (see \reffig{innerProductMat}). This inner product matrix can be perceived as a \emph{functional map} \cite{OvsjanikovBSBG12}, expressing how eigenfunctions on the original domain are mapped to the simplified domain (see \refsec{functional_map_intro}).

Preserving the spectral properties of an operator can be cast as an optimization problem, minimizing the commutative energy \cite{liu2019spectral}
\begin{align} \label{equ:spectralCoarseningEnergy}
   f(\X) = \|\mR \mM^{\text{-}1} \mL \Phi - \widetilde{\mM}^{\text{-}1}\X \mR \Phi \|^2_{\widetilde{\mM}},
\end{align}
where $\mL$ and $\X$ denote the original and the coarsened operators, $\mM$ and $\widetilde{\mM}$ are the original and the coarsened mass matrices, $\mR$ is the restriction operator restricting functions from the original domain to the coarsened domain, and $\Phi$ are the functions (\eg, eigenfunctions) used to measure the commutativity. 

Intuitively, if the coarsened operator $\X$ preserves the spectral properties of the original operator $\mL$, then given some functions $\Phi$ on the original domain, first applying the original operator $\mM^{\text{-}1} \mL$ and then \revise{restricting the functions} to the coarsened domain via $\mR$ should be the same as first restricting \revise{the functions via $\mR$} and then applying the coarsened operator $ \widetilde{\mM}^{\text{-}1} \X$. In the Appendix C of \cite{liu2019spectral} they show that, when $\Phi$ are eigenfunctions, minimizing the commutative energy also preserves eigenvalues.

\paragraph*{Relationship to \cite{liu2019spectral}} Many differential operators in geometry processing are sparse, symmetric, and positive semidefinite. Thus, the method of \cite{liu2019spectral} adds constraints to \refequ{spectralCoarseningEnergy} in order to preserve the three operator properties. They satisfy the constraints via change of variables from $\X$ to $\mG$
\begin{align} \label{equ:nonconvexEnergyLiu2019}
	\minimize_\X\ f(\X) \ \Rightarrow \  \minimize_\mG\ f(\mG^\top \mL \mG).
\end{align}
\revise{where $\mG$ has a predetermined sparsity pattern.} However, this transforms the original convex formulation into a non-convex \revise{\emph{quartic}} one (see Eq.7 in \cite{liu2019spectral}) and increases the sparsity of the output operator to 3-rings. It also artificially limits the feasible region to a subset of PSD matrices determined by $\mG$. In contrast, we will show how to directly optimize \revise{the commutative energy} with respect to $\X$ while maintaining the convexity and \revise{enabling} one to control over the output sparsity.

\subsection{Chordal Spectral Coarsening}\label{sec:chordalSpectralCoarsening}
Spectral coarsening can be written as the following optimization
\begin{alignat}{2}
    &\minimize_{\X}\quad  &&f(\X) \label{equ:opt1} \\ 
    & \text{ subject to}\quad  &&\X \vv = \ve \\
    & && \X \succeq 0 \label{equ:opt1_PSD} \\
    & && \X \in \S^n_{\E}, 
\end{alignat}
where $f$ is the spectral coarsening energy in \refequ{spectralCoarseningEnergy}, $\X \succeq 0$ denotes the PSD constraint, and $\S^n_\E$ denotes \revise{the set of} $n$-by-$n$ sparse symmetric matrices with a user-defined (non-chordal) sparsity pattern $\E$. 
The equality $\X \vv = \ve$ represents the null-space constraint of a differential operator, in the case of Laplacian $\vv = \textbf{1}$ is a constant function and $\ve = \textbf{0}$ \revise{is a zero vector} \revise{because every row or column of a Laplacian sums to zero}.
For the sake of simplicity, we describe the entire process without expanding the spectral coarsening energy $f$, and the complete formulation is detailed in \refapp{argminX_spectralCoarsening}.

Applying the chordal extension (\refsec{chordalExtension}) and the chordal decomposition (\refsec{chordalDecomposition}) to \refequ{opt1} leads to
\begin{alignat}{3}
    &\minimize_{\X,\{\Z_\i\}}\quad  &&f(\X) \label{equ:opt2} \\
    & \text{ subject to}\quad  &&\X \vv = \ve \label{equ:opt2_null}\\
    & && \X \in \S^n_\C  \\
    & && \X_{jk} = 0, \quad &&\forall (j,k) \in \C \backslash \E \\
    & && \X = \sum_{\i=1}^p \P_\i^\top \Z_\i \P_\i \label{equ:opt2_chordal}\\
    & && \Z_\i \succeq 0, \quad &&k = 1,\cdots, p, \label{equ:opt2_psd} 
\end{alignat}
where $p$ is the number of maximal cliques. We convert the PSD constraint $\X \succeq 0$ in \refequ{opt1_PSD} to many small PSD constraints $\Z_\i \succeq 0$ according to \refthm{chordalPSD}. Here we also perform chordal extension to switch the sparsity from non-chordal $\E$ to a chordal $\C$ with additional equality constraints $\X_{jk} = 0$ (see \refequ{chordalExtension}). 

Ensuring the PSD property of the matrix requires a full (generalized) eigendecomposition followed by the removal of the negative eigenvalues.
When the matrix is large, a full decomposition is intractable to compute. Using chordal decomposition to transform the big PSD constraint \revise{(}\refequ{opt1_PSD}\revise{)} to a set of small ones \revise{(}\refequ{opt2_psd}\revise{)} allows us to efficiently project each $\Z_\i$ to PSD in parallel.


\subsection{Change of Variables}
Utilizing the fact that $\X$ is symmetric with a sparsity pattern $\E$, we propose to accelerate the solver via change of variables from $\X$ to a compressed vector $\xE$ which \revise{consists of} the non-zero elements of the lower triangular part defined by $\E$.
This change of variables restricts the optimization to search only within the feasible sparsity $\E$.
This is crucial to the performance of the solver because $\X$ is sparse thus $|\xE| \ll |\text{vec}(\X)|$ significantly reduces the degrees of freedom. The relationship between $\X$ and $\xE$ is described by
\begin{align}\label{equ:x2xE}
    \text{vec}(\X) = \invP_\E \xE, \qquad \xE = \P_\E \text{vec}(\X),
\end{align} 
\begin{wrapfigure}[5]{r}{1.33in}
	\raggedleft 
    \vspace{-15pt}
	\hspace*{-0.7\columnsep}
	\includegraphics[width=1.31in, trim={6mm 0mm -1mm 0mm}]{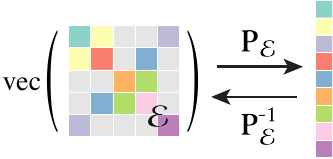} 
	\label{fig:vectorizeX}
\end{wrapfigure} 
where $\P_\E$ is a selection matrix to the sub-vector $\xE$. $\invP_\E$ is the inverse of $\P_\E$ which is another matrix to re-index elements in $\xE$ back to $\text{vec}(\X)$. Note that $\invP_\E$ is different from the $\P_\E^\top$ as each non-diagonal element in $\xE$ gets mapped to two entries in $\text{vec}(\X)$, instead of one entry, and $\invP_\E$ can be assembled easily without the need of explicitly inverting the matrix. This change of variables incorporates both the chordal symmetric constraint $\X \in \S^n_\C$ and the equality constraints $\X_{jk} = 0$ in \refequ{opt2}. After some derivation in \refapp{changeOfVariables}, we have
\begin{alignat}{3}
    &\minimize_{\xE,\{\z_\i \}}\quad  &&f(\xE) \label{equ:opt3} \\
    & \text{ subject to}\quad  &&\mG \xE = \ve \\
    & && \invP_\E \xE = \sum_{\i = 1}^p \K_\i \z_\i \label{equ:opt3_vecChordal}\\
    & && \text{vec}^{\text{-}1}(\z_\i) \succeq 0, \quad &&\i = 1,\cdots, p, 
\end{alignat}
We define $\z_\i \coloneqq \text{vec}(\Z_\i)$ to be the vectorized clique matrix. $\mG \xE = \ve$ is the vectorized version of the $\X \vv = \ve$ in \refequ{opt2_null}. $\invP_\E \xE = \sum_{\i = 1}^p \K_\i \z_\i$ is the vectorized chordal decomposition \refequ{opt2_chordal}. Here $\K_\i$ denotes the index selection matrix for vectorized clique matrix $\z_\i$.

We use another change of variables to further accelerate the algorithm by restricting the vectorized chordal decomposition in \refequ{opt3_vecChordal} to only the non-zeros in the chordal sparsity pattern $\C$. That is because the summation of $\{\z_\i\}$ in \refequ{opt3_vecChordal} only has non-zeros in the chordal sparsity pattern $\C$. We introduce another index selection matrix $\P_\C$ to change \refequ{opt3_vecChordal} into
\begin{align}\label{equ:reduceToC}
    \invP_\E \xE = \sum_{\i = 1}^p \K_\i \z_\i\ \Rightarrow\ \P_\C \invP_\E \xE = \P_\C \sum_{\i = 1}^p \K_\i \z_\i,
\end{align}
where $\P_\C$ selects the lower triangular non-zeros in $\C$ from the original $\text{vec}(\X)$. Here $\P_\C$ is defined the same as the $\P_\E$ in \refequ{x2xE} but with a different sparsity pattern $\C$.

\begin{wrapfigure}[4]{r}{1.33in}
    \raggedleft
    \vspace{-20pt} 
	\hspace*{-0.7\columnsep}
	\includegraphics[width=1.31in, trim={6mm 0mm 1mm 0mm}]{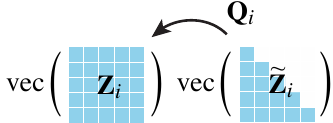} 
	\label{fig:selectTril}
\end{wrapfigure} 
As $\z_\i$ is the vectorization of a \emph{symmetric} matrix $\Z_\i$, another reduction is achieved \revise{by} applying the same trick as \refequ{x2xE} to restrict the degrees of freedom of $\Z_\i$ to \revise{its} lower triangular part $\rZ_\i$ via an expansion matrix $\Q_\i$ (see the inset).
\begin{align}
    \underbrace{\text{vec}(\Z_\i)}_{\z_\i} = \Q_\i \underbrace{\text{vec}(\rZ_\i)}_{\rz_\i},
\end{align}
We use $\z_\i, \rz_\i$ to denote the vectorized $\Z_\i$ and the vectorized lower triangular part $\rZ_\i$, respectively. 
We define $\Q_\i$ as an inverse index selection matrix that expands the vector of the lower triangular element $\rz_\i$ to $\z_\i$. 

Combining the above results leads us to the reduced optimization problem
\begin{alignat}{3}
    &\minimize_{\xE,\rz}\quad  &&f(\xE)\label{equ:reduce_opt3} \\
    & \text{ subject to}\quad  &&\mG \xE = \ve \label{equ:null_space_constraint}\\
    & && \P_\C \invP_\E \xE = \P_\C \rK \rz \\
    & && \text{vec}^{\text{-}1}( \Q_\i \rz_\i)  \succeq 0, \quad &&\i = 1,\cdots, p,
\end{alignat}
where
\begin{align}
    \rK = \begin{bmatrix}
        \K_1 \Q_1, \ 
        \cdots, \ 
        \K_p \Q_p
    \end{bmatrix},\quad
    \rz = \begin{bmatrix}
        \rz_1 \\
        \vdots \\
        \rz_p
    \end{bmatrix}.
\end{align}
This final reduced formulation is an optimization problem \revise{which involves} only linear equalities and small dense PSD constraints. We solve this optimization using ADMM \revise{(see \refapp{ADMM})}, alternating between solving for $\xE$ and $\rz$. Solving for $\xE$ when $f$ is the spectral coarsening energy boils down to a single linear solve; solving for $\rz$ leads to a subroutine of projecting each clique matrix to PSD by removing the negative eigenvalues. The update on $\rz$ is efficient as each $\rz_\i$ is small and can be trivially parallelized. We provide details of the ADMM derivation in \refapp{admmForm}.

\subsection{Weighted Spectral Coarsening}
Solving \refequ{reduce_opt3} results in a coarsened operator that preserves the spectral properties of the original one. One can freely control the sparsity pattern of the output by changing $\E$. In our experiments, we choose either 1-, 2-, or 3-ring sparsities. The more rings in use, the better the results because we have more degrees of freedom in minimizing the spectral coarsening energy \refequ{spectralCoarseningEnergy}. 

When the degrees of freedom are limited, such as using only 1-ring, we notice that the solver would emphasize preserving relatively higher frequencies and lead to worse performance in preserving the lowest frequencies. In response, we weight the spectral coarsening energy \refequ{spectralCoarseningEnergy} via the inverse of eigenvalues, \revise{which} leads to this weighted version
\begin{align} \label{equ:weightedSpectralCoarseningEnergy}
    f_w(\X) = \|\mR \mM^{\text{-}1} \mL \Phi \Lambda^{\text{-}1} - \widetilde{\mM}^{\text{-}1}\X \mR \Phi \Lambda^{\text{-}1} \|^2_{\widetilde{\mM}},
\end{align}
where $\Lambda$ is a diagonal matrix of the eigenvalues of the original operators $\mL$. In \refsec{results}, we show that the weighted version leads to a better spectral preservation in the low frequencies when using our solver. 
This weighted formulation also naturally captures the notion of ``null-space reproduction'' in \refequ{null_space_constraint}, as we explicitly enforce the null-space corresponding to the eigenvalue $0$ as a hard constraint, \ie, with infinite weight.

%

\section{Results} \label{sec:results}
We evaluate our solver by comparing against the existing state-of-the-art spectral coarsening \cite{liu2019spectral} and simplification \cite{lescoat2020spectral},
%
using functional maps and the quantitative metrics $\Vert \cdot \Vert_L$ and $\Vert \cdot \Vert_D$ proposed in \cite{lescoat2020spectral}. 
%
We further demonstrate the power of our solver in controlling the sparsity patterns, approximating volumetric behavior using only boundary surface vertices and detaching the differential operator from the mesh. We provide implementation details in \refapp{implementation}.

\revise{
\subsection{Evaluation Metrics} \label{sec:functional_map_intro}
Functional maps \cite{OvsjanikovBSBG12} describe how to transport functions from one shape $\mathcal{M}$ to another shape $\mathcal{N}$. 
The idea of functional map \revise{has} led to breakthroughs in computing shape correspondences \cite{OvsjanikovCBRBG17}. 
In the context of spectral coarsening, functional maps become a tool for evaluating how the eigenvectors of a discrete operator $\mL \in \R^{n \times n}$ derived on a high-resolution mesh are maintained by a coarsened operator $\X \in \R^{m \times m}$. Following the notation in \reffig{innerProductMat}, let $\Phi \in \R^{n \times k}$ and $\widetilde{\Phi} \in \R^{m \times k}$ be two set of eigenvectors of $\mL$ and $\X$, respectively, the functional map $\mC$ can be computed as  
\begin{align}
    \mC = \widetilde{\Phi}^\top \widetilde{\mM} \mR \Phi.
\end{align}
\begin{wrapfigure}[5]{r}{0.16\linewidth}
    \raggedleft
    \vspace{-54pt} 
	\hspace*{-0.6\columnsep}
    \begin{minipage}[b]{\linewidth}
    \includegraphics[width=0.5in, trim={4mm 20mm 0mm 20mm}]{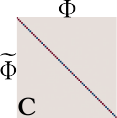}
    \label{fig:perfectFunctionalMap}
    \end{minipage}
  \end{wrapfigure}
Here $\widetilde{\mM}$ is the mass matrix in the coarse domain and $\mR$ is a restriction operator, encoding the correspondences information from the original mesh to its coarsened counterpart. 
The restriction operator is computed either during the decimation \cite{lescoat2020spectral} or simply a subset selection matrix as in \cite{liu2019spectral}. 
One can also perceive the matrix $\mC$ as an inner product matrix between the eigenvectors $\widetilde{\Phi}$ on the coarsened domain and the restricted eigenvectors $\mR \Phi$ to the coarsened domain. 
Due to the orthonormality between eigenvectors, the optimal functional map (or inner product matrix) $\mC$ should be a diagonal matrix of values 1 and -1 (see inset). 

\emph{Laplacian commutativity and Orthonormality norm.}  The functional map should be orthonormal and commute with the original Laplace operator in the reduced basis if and only if it preserves corresponding eigenfunctions and eigenvalues exactly, as shown in \cite{lescoat2020spectral}. Thus the spectral preservation before and after coarsening and simplification can be quantified using two norms:
\begin{align}
    \text{Laplacian commutativity: } &\Vert \cdot \Vert_L^2 = \frac{\Vert \mC \Lambda - \widetilde{\Lambda} \mC\Vert^2}{\Vert \mC \Vert^2} \\ 
    \text{Orthonormality: } &\Vert \cdot \Vert_D^2 = \Vert \mC^\top \mC - \mI \Vert^2.
\end{align}
%

In our experiments, we visualize the functional map $\mC$ and report both norms to convey a complete picture of spectral preservation.
}

\begin{figure}
    \centering
    \includegraphics[width=3.33in]{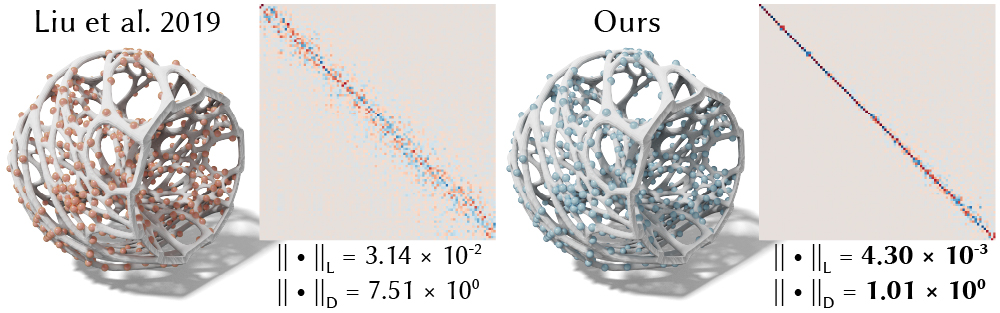}
    \caption{Using the same 3-ring sparsity pattern, our convex formulation enables the ADMM solver to converge to a better result on shape (from 80,000 vertices to 600) where the gradient descent in \cite{liu2019spectral} may struggle to converge.
    }
    \label{fig:spec_coarse_complicated_cell}
    \vspace{-10pt}
\end{figure} 
\begin{figure}
    \centering
    \includegraphics[width=3.33in]{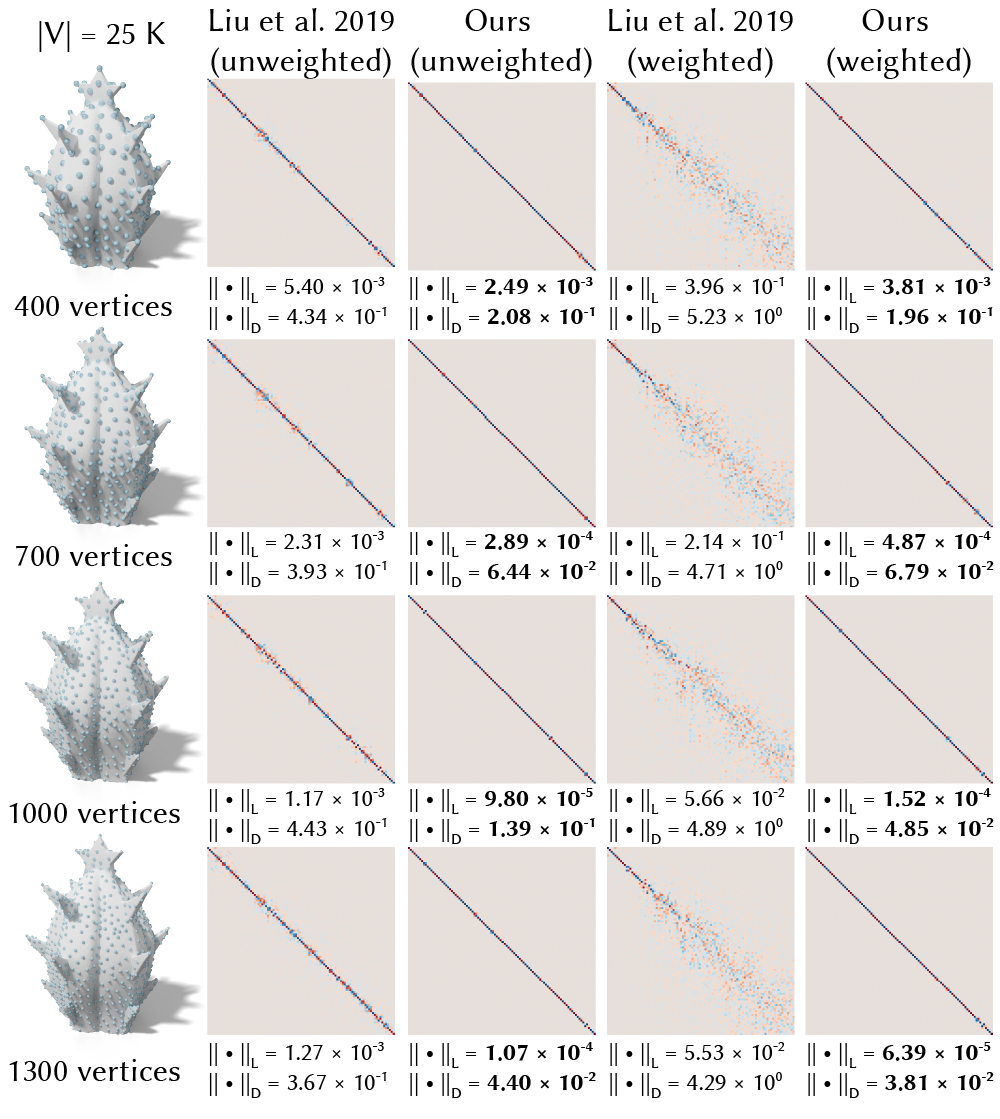}
    \caption{Using the same 3-ring sparsity as \cite{liu2019spectral}, our method achieves better quality of resulting functional maps for both the weighted and unweighted versions, measured by the metrics proposed in \cite{lescoat2020spectral}.}
    %
    \label{fig:spec_coarse_compare_cactus_25K}
    \vspace{-10pt}
\end{figure} 
\begin{figure}
    \centering
    \includegraphics[width=3.33in]{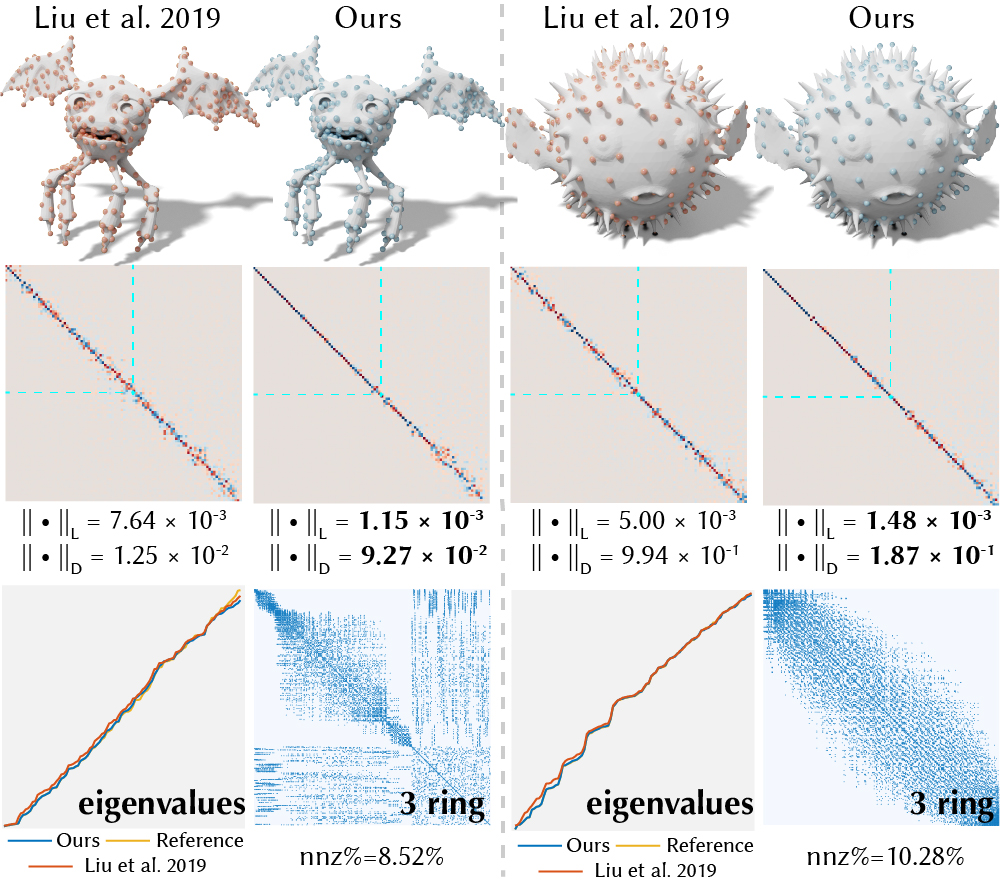}
    \caption{\revise{For applications that desire to preserve low frequencies, our weighted formulation can focus on preserving the first few eigenvectors and eigenvalues (shown in increasing order).
    Our weighted formulation achieves better results comparing to \cite{liu2019spectral} under the same sparsity pattern when coarsening the shapes from 8,000 (left) and 28,000 (right) vertices to 400, respectively.}
    Here we show the Laplacian commutativity norm and Orthonormality norm based on the functional map of the first 50 eigenvectors (inside the dashed lines).
    }
    %
    \label{fig:spec_coarse_complicated_shape}
    \vspace{-0pt}
\end{figure} 
\begin{figure}
    \centering
    \includegraphics[width=3.33in]{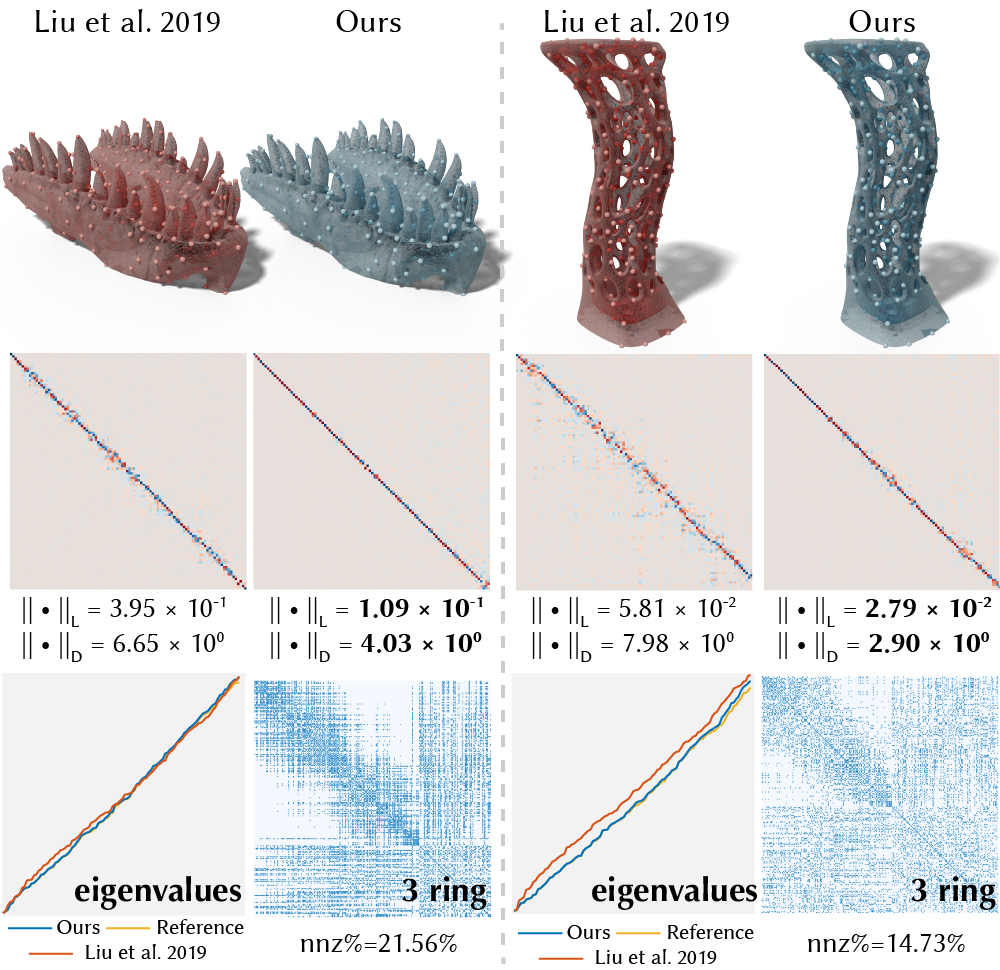}
    \caption{As degrees of freedom increase for volumetric Laplacian, our method is still able to maintain the spectral properties of the tetrahedral meshes (from 32,000 and 27,000 vertices to 400 respectively) using the same sparsity as \cite{liu2019spectral}. 
    \revise{Here the eigenvalues are shown in increasing order.}
    }
    \label{fig:spec_coarse_tetmesh}
    \vspace{-0pt}
\end{figure} 
\subsection{Spectral Coarsening}
\begin{figure}
    \centering
    \includegraphics[width=3.33in]{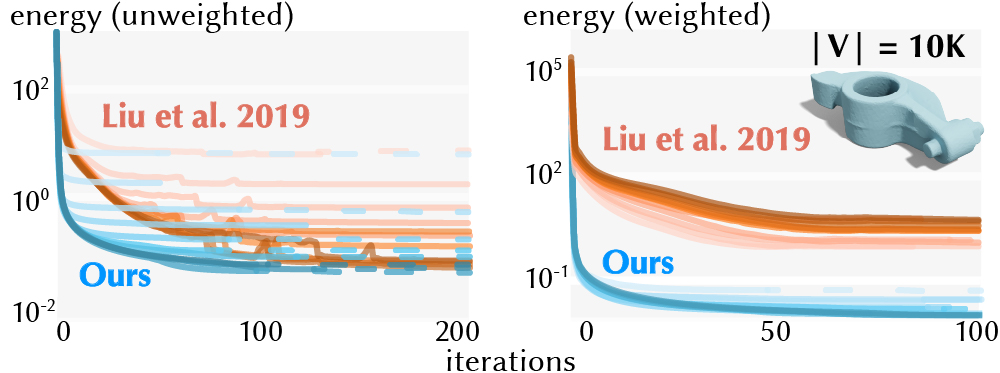}
    \caption{
        Under the same 3-ring sparsity, our method consistently achieves better objective values comparing to the spectral coarsening method proposed by \cite{liu2019spectral}. We evaluate both unweighted (top) and the weighted (bottom) versions across different numbers of coarse vertices ranging from 200 (light) to 1200 (dark). Note that the dashed lines denote that the optimization has already converged.}
    \label{fig:spec_coarse_compare_energy_both}
    \vspace{-5pt}
\end{figure} 
Comparing to the original non-convex formulation \refequ{nonconvexEnergyLiu2019} \cite{liu2019spectral}, in \reffig{spec_coarse_compare_energy_both} we show that our convex formulation consistently achieves lower objective values across different number of coarsened vertices (from 200 to 1200) and leads to better qualitative results (see \reffig{spec_coarse_compare_cactus_25K}, \revise{\reffig{spec_coarse_complicated_cell} and \reffig{spec_coarse_tetmesh}}).
For a fair comparison, we set the sparsity pattern of our approach to be 3 rings, the same as the method of \cite{liu2019spectral}. 
We can further show that through weighting the energy with the inverse of the eigenvalues (see \refequ{weightedSpectralCoarseningEnergy}), we obtain \revise{an} even better preservation of the low-frequencies, see \reffig{spec_coarse_compare_cactus_25K} (right two) and \revise{\reffig{spec_coarse_complicated_shape}}.
In general, the weighted version performs better in maintaining the lowest frequencies, while the unweighted version tends to preserve all the eigenmodes in a least-square sense.
\begin{figure}
    \centering
    \includegraphics[width=3.33in]{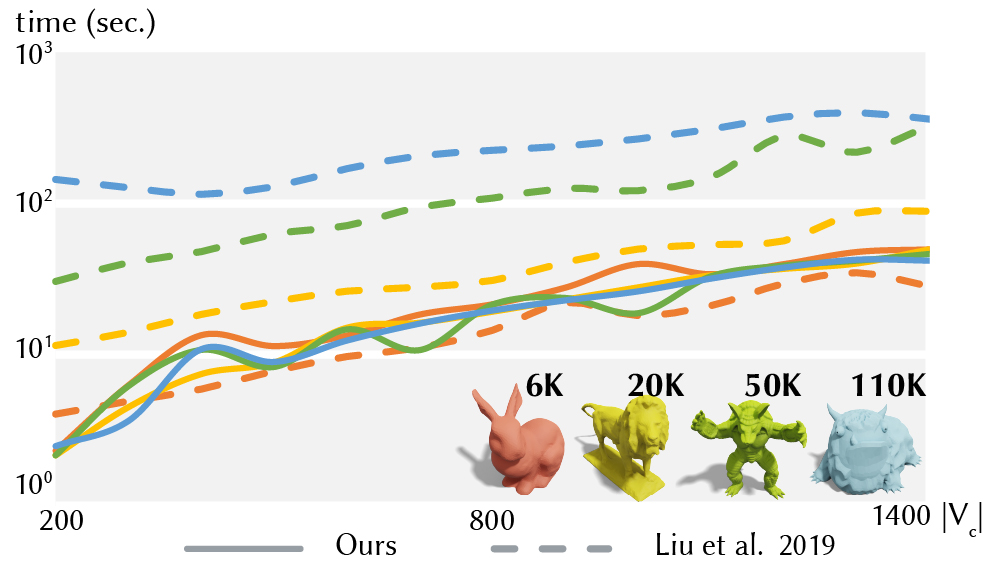}
    \caption{We compare the runtime of our algorithm (weighted) with \cite{liu2019spectral} (unweighted) using the same 3-ring sparsity pattern with respect to the number of coarsened vertices $|\textsc{V}_c|$, as our method performs better with the weighted version and \cite{liu2019spectral} shows the opposite.
    Here we only consider the solve time, factoring out the precomputation for both our method and the method of \cite{liu2019spectral}.
    %
    %
    As in our formulation the solve involved in ADMM is independent of the resolution of the original mesh, we are able to coarsen a high-resolution mesh without a significantly increased solve time compared to \cite{liu2019spectral}.
    }
    \label{fig:spec_coarse_compare_best_time}
    \vspace{-5pt}
\end{figure}

\begin{figure}
    \centering
    \includegraphics[width=3.33in]{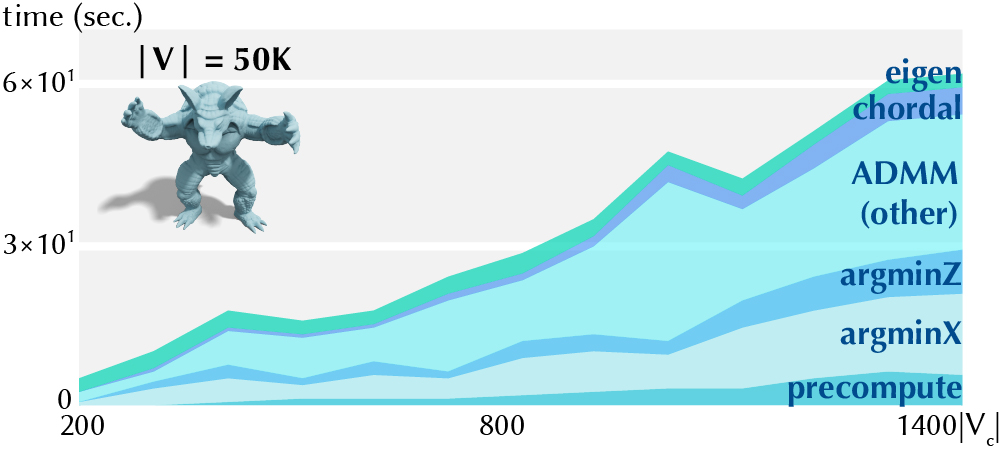}
    \caption{We show the decomposition of the total runtime of our algorithm using the same 3 rings sparsity pattern as in \reffig{spec_coarse_compare_best_time}.
    From bottom to top are \revise{the precomputation time,} $\argmin_\X$ time, $\argmin_\Z$ time, other ADMM time (including numerical factorization), chordal decomposition time and \revise{eigendecomposition time}, respectively. 
    As shown in the figure, most of the runtime of our algorithm is spent on numerical factorization. 
    By reusing numerical factorization until $\rho$ changes, the time spent on each $\argmin_\X$ and $\argmin_\Z$ step is relatively small.
    }
    \label{fig:spec_coarse_runtime_decomposition_armadillo}
    \vspace{-5pt}
\end{figure} 

\begin{figure}
    \centering
    \includegraphics[width=3.33in]{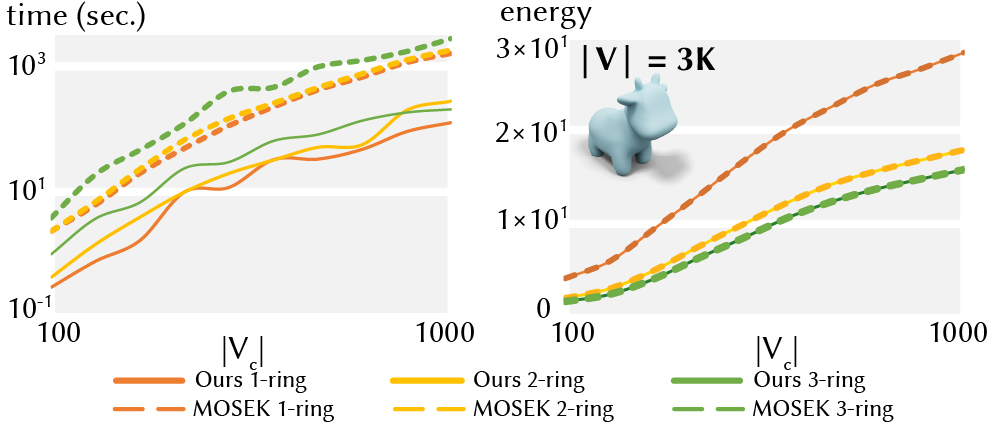}
    \caption{We compare the total runtime of our solver and the MOSEK solver in CVX \cite{cvx, cvx_gb08}, which \revise{only supports dense SDP constraints and} uses interior point method to solve the dense SDP problem using the 1-, 2- and 3-ring sparsity patterns of \cite{GarlandH97}. 
    As a dense \revise{SDP} solver that is not designed to solve large sparse SDP problem, MOSEK takes a relatively long runtime when the matrix size is large or the rings of neighborhood increases.
    Here $|\textsc{V}_c|$ is number of the vertices in the coarse mesh.
    }
    \label{fig:mosek_runtime_energy_comparison}
    \vspace{-5pt}
\end{figure}

%
With the reusable numerical factorization and separable PSD projection structures, our ADMM solver is able to solve the problem efficiently while the method of \cite{liu2019spectral} takes longer to converge.
In \reffig{spec_coarse_compare_best_time}, we compare the runtime with \cite{liu2019spectral}, both using the optimal setups (our weighted version and \cite{liu2019spectral} unweighted version).
The $\argmin_\X$ step requires a linear solve of a KKT system and $\argmin_\Z$ are a set of PSD projections of the small clique matrices. \revise{For details about $\argmin_\X$ and $\argmin_\Z$ step, see \refapp{ADMM}. }
Leveraging the fact that the KKT system matrix in $\argmin_\X$ remains the same until $\rho$ is updated, we only perform numerical factorization when $\rho$ is updated and reuse it until $\rho$ changes again (usually after tens of iterations). 
As shown in \reffig{spec_coarse_runtime_decomposition_armadillo}, most of our runtime is spent on numerical factorization while the time spent on each $\argmin_\X$ and  $\argmin_\Z$ step is relatively small.
\revise{We report our detailed runtime in \reffig{spec_coarse_runtime_eigen}. }
For detailed runtime comparison within the weighted and unweighted versions, see \reffig{spec_coarse_compare_time_armadillo_both}.

We also compare the total runtime of our sparse ADMM solver with the MOSEK solver in CVX \cite{cvx,cvx_gb08} in \reffig{mosek_runtime_energy_comparison}, which uses the interior point method to solve the problem with dense PSD constraints. 
We show our solver can \revise{work} on large problems in \revise{a} more efficient way than MOSEK, while MOSEK, \revise{which only supports dense semi-definiteness constraints, is not designed for large sparse SDP problem and} takes a relatively long time to converge when the matrix size is large or the rings of neighborhood increases.
Here we use $ 0.8 \times |\textsc{V}_c|$ eigenvectors to ensure both methods converge.

\subsection{Spectral Simplification}
\begin{figure}
    \centering
    \includegraphics[width=3.33in]{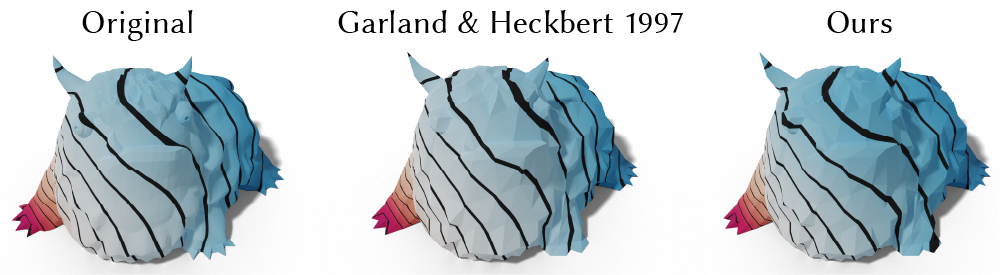}
    \caption{We visualize the biharmonic distance of our method and \cite{GarlandH97} using the same 1-ring sparsity pattern. Our method can further postprocess and improve spectral preservation of the result from \cite{GarlandH97} (from 110,000 vertices to 500).}
    %
    \label{fig:qslim_biharmonic_monster}
    \vspace{-5pt}
\end{figure} 
\begin{figure}
    \centering
    \includegraphics[width=3.33in]{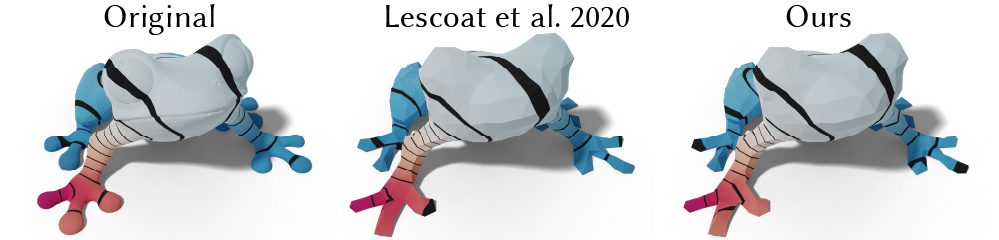}
    \caption{By visualizing the biharmonic distance, we show that our approach can also postprocess the result from \cite{lescoat2020spectral} (from 10,000 vertices to 800) and achieve better spectral preservation while still maintaining the same 1-ring sparsity pattern.}
    %
    \label{fig:spec_simpl_biharmonic_frog}
    \vspace{-3pt}
\end{figure} 
\begin{figure}
    \centering
    \includegraphics[width=3.33in]{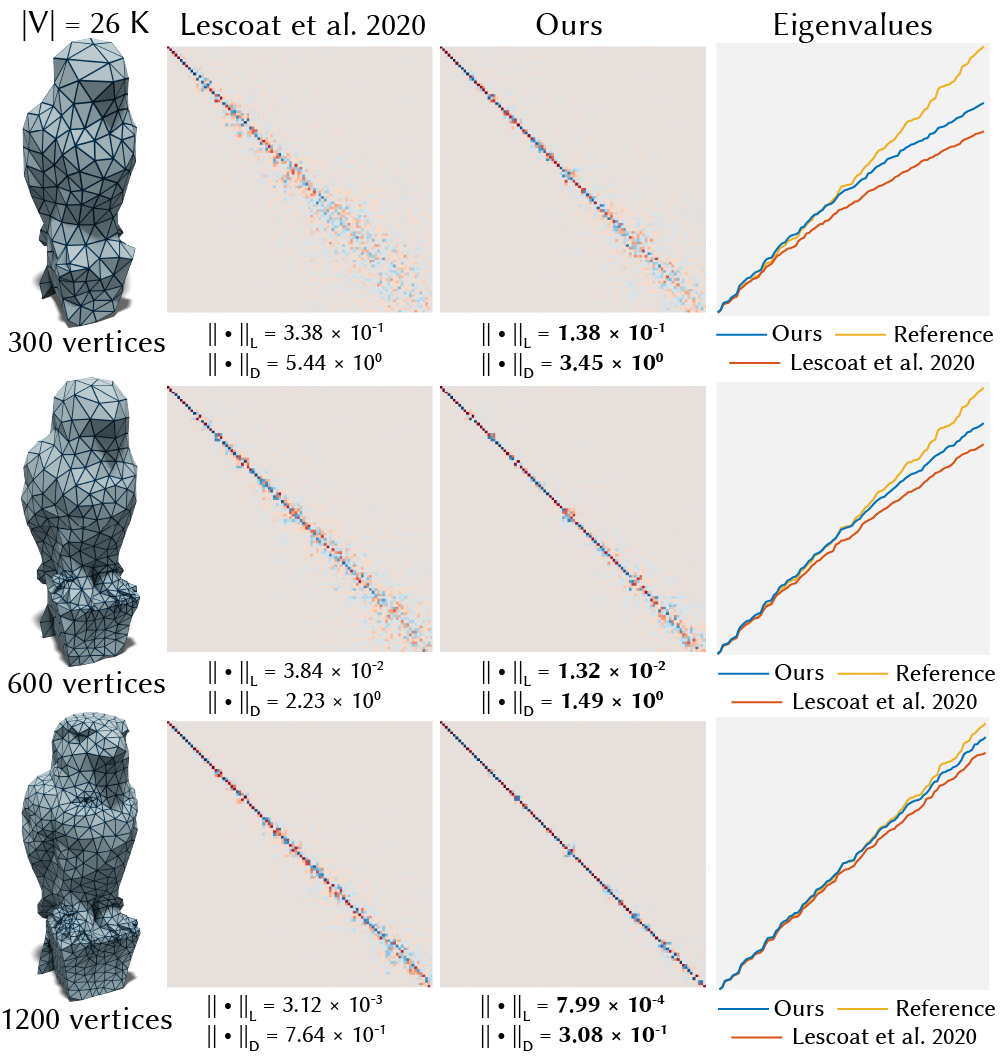}
    \caption{Using the same 1-ring sparsity pattern, our method can serve as a post-processing tool to further improve the resulting operator from the method of \cite{lescoat2020spectral}.
    Given the original mesh with 26,000 vertices, our post-processed operators result in better functional maps (middle) compared to the output operators from \cite{lescoat2020spectral} (left), as well as closer eigenvalues (right) to the reference.}
    \label{fig:spec_simpl_compare_eagle_26K}
    \vspace{-10pt}
\end{figure} 
\begin{figure}
    \centering
    \includegraphics[width=3.33in]{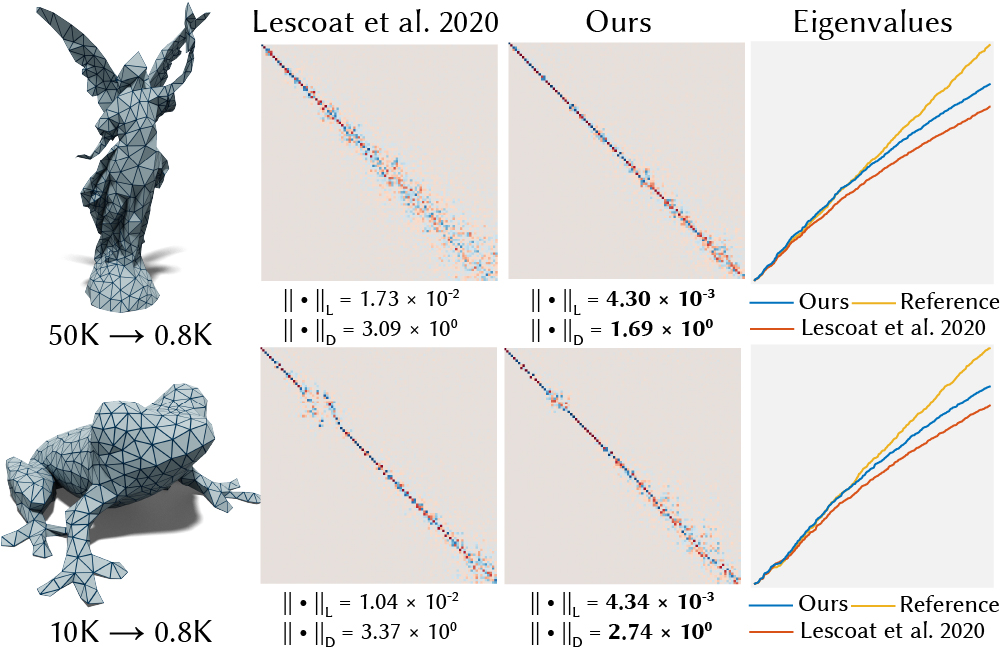}
    \caption{When the coarsening is aggressive, our method can still postprocess the results of \cite{lescoat2020spectral} to improve the quality of spectral preservation.
    }
    \label{fig:spec_simpl_refine_lucy_frog}
    \vspace{-10pt}
\end{figure} 
\begin{figure}
    \centering
    \includegraphics[width=3.33in]{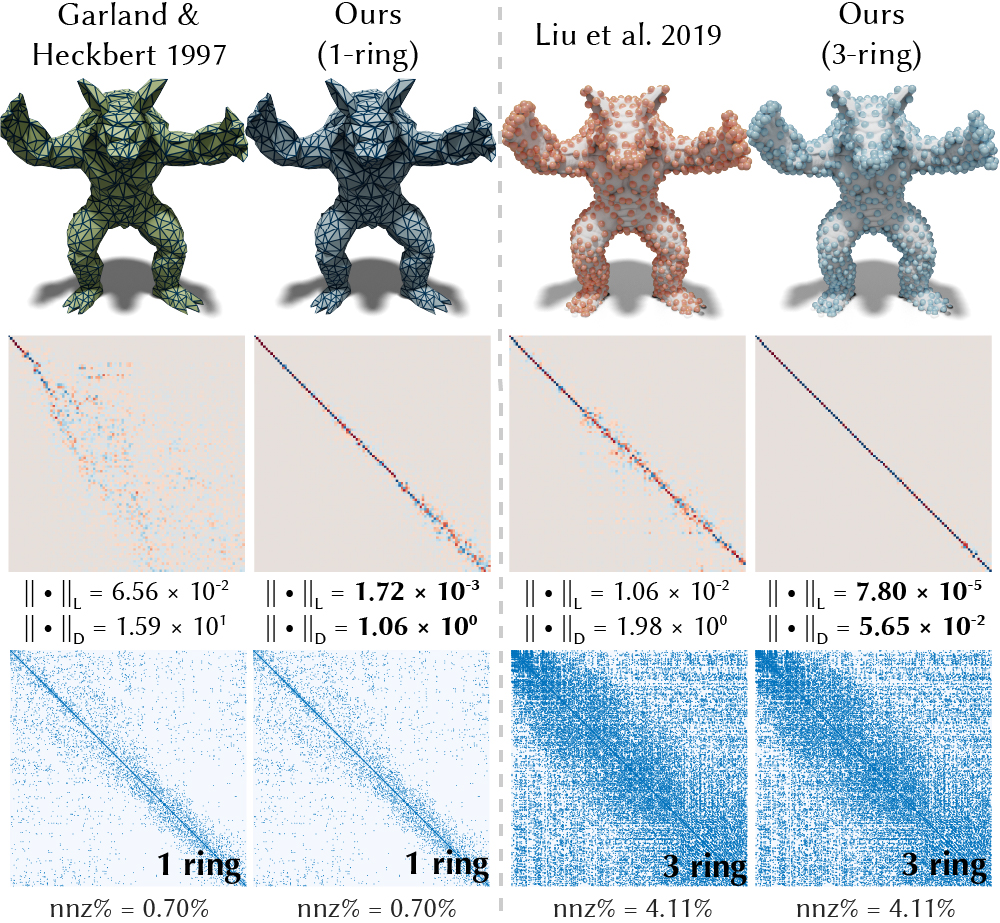}
    \caption{We simplify the anisotropic Laplacian (with parameter 20) from 50,000 vertices to 1,000 vertices using the same sparsity pattern as \cite{GarlandH97} or \cite{liu2019spectral}. Our method can handle anisotropic operators where \cite{GarlandH97} may fail entirely due to the \revise{anisotropy}. Our optimization scheme enables users to freely choose between 1-ring or 3-ring sparsity. In contrast, \cite{liu2019spectral} has much less control on the sparsity pattern and only allows for 3-ring sparsity pattern, which introduces a significant amount of fill-ins. 
    }
    \label{fig:anisotropic_armadillo}
    \vspace{-0pt}
\end{figure} 
\begin{figure}
    \centering
    \includegraphics[width=3.33in]{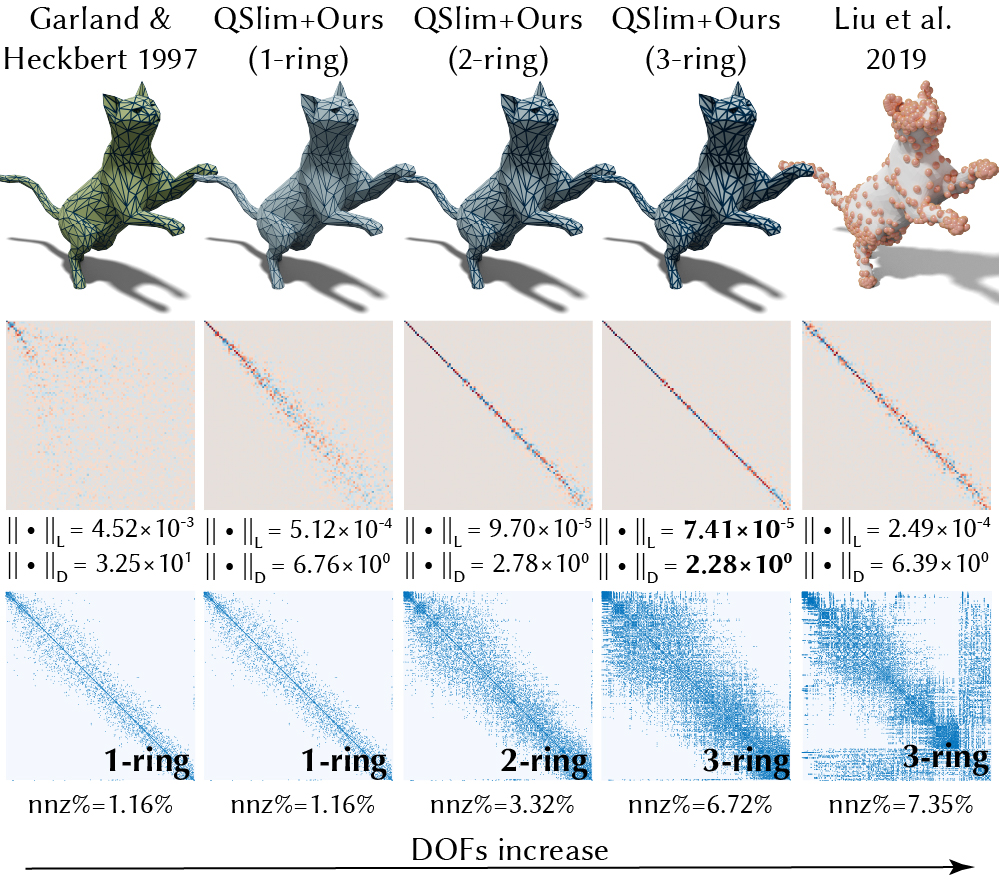}
    \caption{Our optimization achieves better spectral preservation of the anisotropic Laplace operator (with parameter 60, from 5000 vertices to 400 vertices) when the rings of neighborhood increases. Increasing the non-zeros in the sparsity pattern will allow for more degrees of freedom, which enables our solver to converge to a better result. 
    }
    \label{fig:anisotropic_123ring_TOSCA_cat_sf3}
    \vspace{-0pt}
\end{figure} 
Our approach could further improve the results from the spectral simplification via post-processing.
The method of \cite{lescoat2020spectral} performs spectral simplification by greedily collapsing the edge with the minimum cost, thus it may result in suboptimal results. 
In \reffig{spec_simpl_compare_eagle_26K} and \reffig{spec_simpl_compare_monster_110K} we post-process the cotangent Laplacian from the results of \cite{lescoat2020spectral} in a global manner to further improve the spectral preservation while keeping the sparsity pattern and the mesh vertices fixed.
\revise{We further demonstrate the improvement of the spectral preservation by visualizing the biharmonic distance of our method and \cite{GarlandH97} (\reffig{qslim_biharmonic_monster}) or \cite{lescoat2020spectral} (\reffig{spec_simpl_biharmonic_frog}) using the same 1-ring sparsity pattern. }
Our method can also recover the spectral properties when the coarsening is extreme for complicated shapes (see \reffig{spec_simpl_refine_lucy_frog} and \reffig{spec_simpl_refine_chess_cupman}).
In addition to the isotropic cotangent Laplacian, in \reffig{anisotropic_armadillo} we demonstrate our capability in handling anisotropic operators without introducing any new fill-ins.
%

For downstream applications that accept changes in the sparsity pattern, our method enables one to freely control the sparsity patterns to achieve better results.
As shown in \reffig{anisotropic_123ring_TOSCA_cat_sf3}, we can freely increase the sparsity pattern from 1 ring to 3 rings in order to allow more degrees of freedom and better results.
But one should also consider the trade-off between the number of non-zero fill-ins and the quality of the results because more degrees of freedom implies a denser output operator with a longer runtime (see \reffig{mosek_runtime_energy_comparison}). 
%

\subsection{Volume to Surface}
\begin{figure}
    \centering
    \includegraphics[width=3.33in]{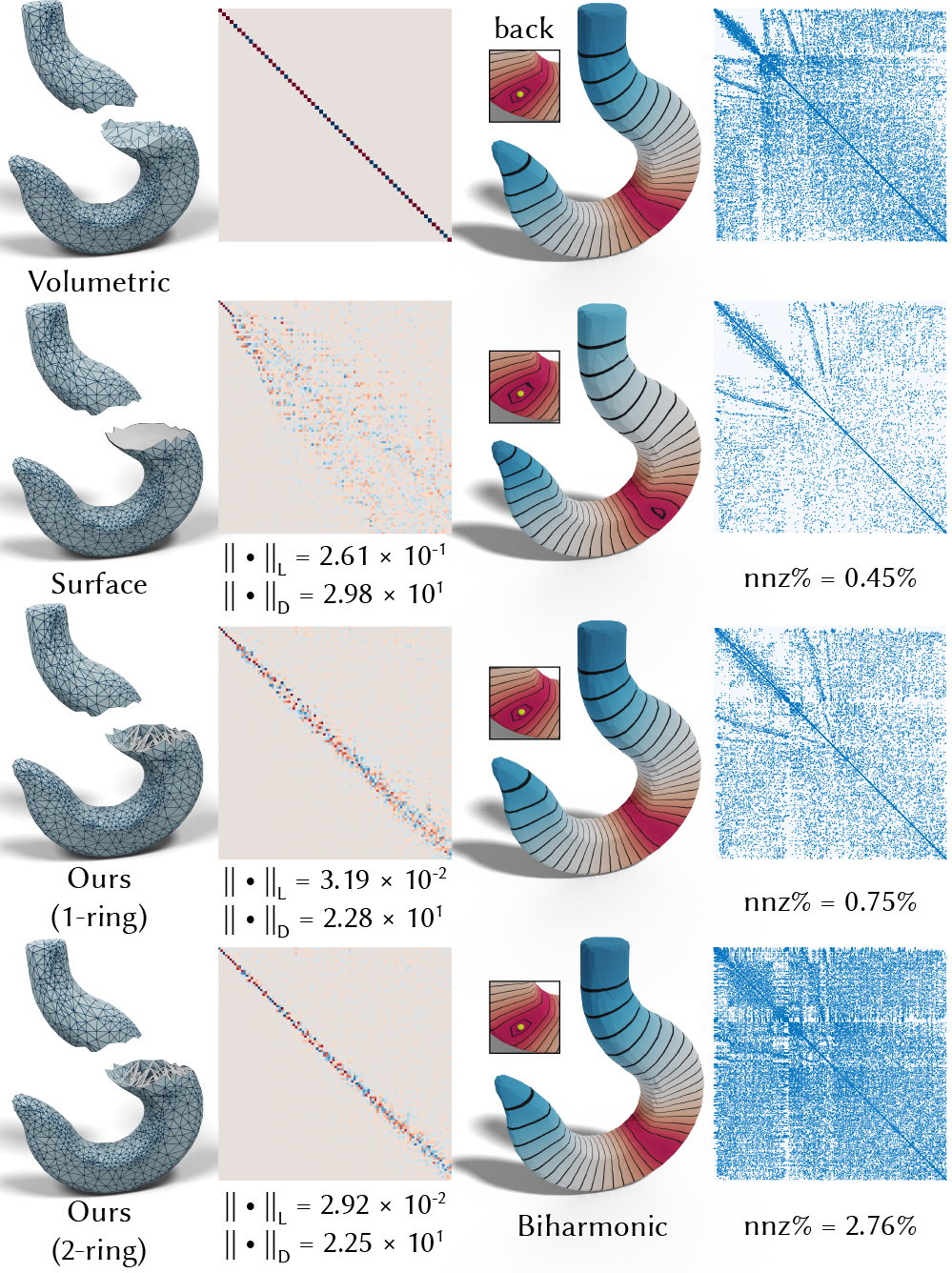}
    \caption{Our method can encode the spectral behavior of a volumetric mesh only using its surface mesh with some added links. 
    We approximate the volumetric behavior using a \emph{sparse} matrix with a controllable sparsity pattern, while the corresponding matrix has to be \emph{dense} in the traditional Boundary Element Method \cite{ArtDefo1999}.
    Here the source vertex of the biharmonic distance is visualized as a green dot, and the added links are visualized as the gray lines (bottom two). 
    }
    \label{fig:vol2surf_craneHook} 
    \vspace{-12pt}
\end{figure} 
\begin{figure}
    \centering
    \includegraphics[width=3.33in]{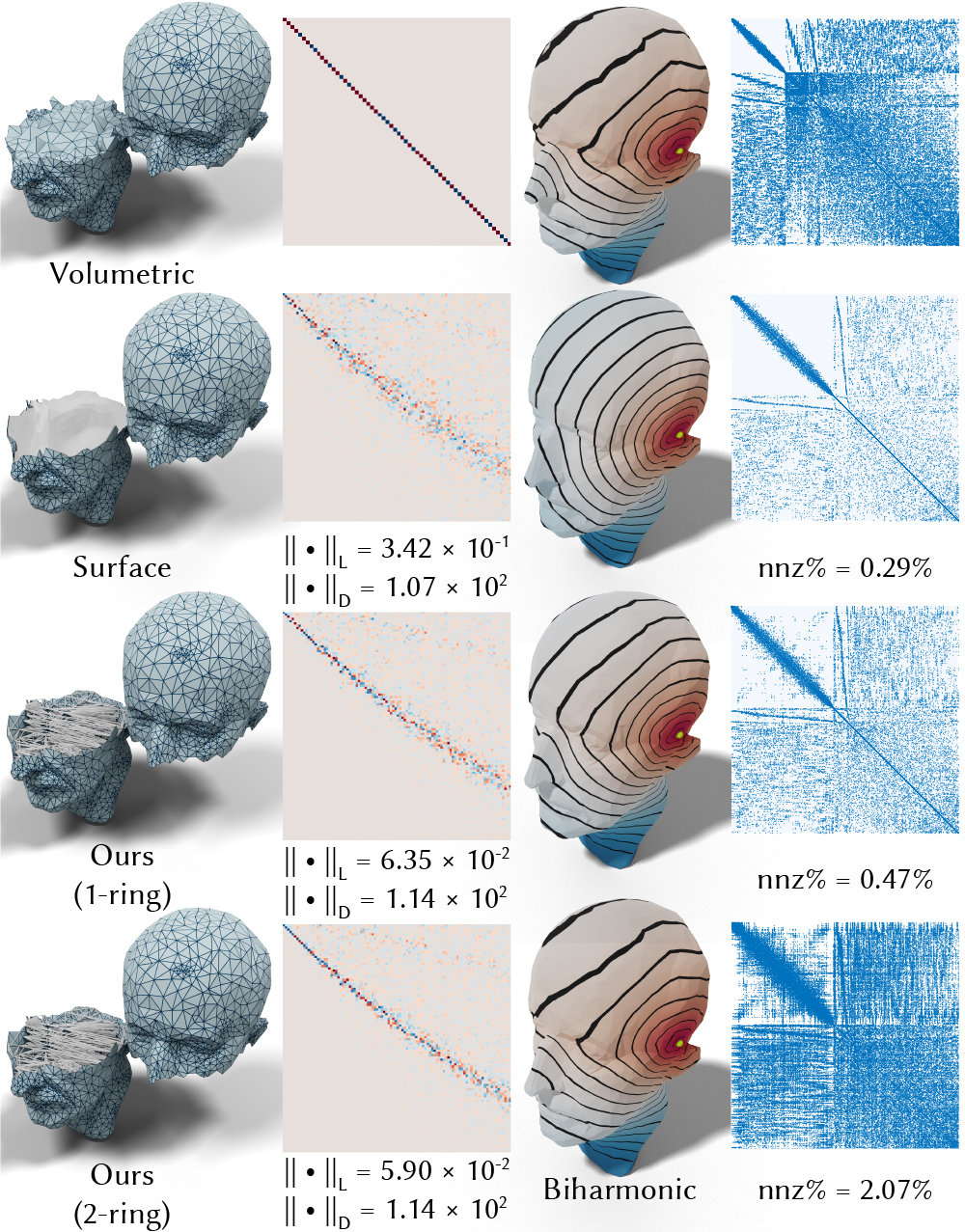}
    \caption{Starting from the constrained Delaunay \revise{tetrahedralization}, we can increase the number of rings of neighborhood to better approximate the volumetric Laplacian using a surface mesh with random links.
    Similar to the partial functional correspondence in \cite{partialFMap2017}, the diagonal of our functional map may be skewed because we may lose some internal eigenvectors during this partial matching. }
    \label{fig:vol2surf_max_planck} 
    \vspace{-12pt}
\end{figure} 
Surface-only representation is a more efficient alternative compared to its volumetric counterpart because three dimensional (volumetric) problem is reduced to two dimensions (surface).
However, in computer animation and simulation, it is often more desirable to use a volumetric representations to simulate the volumetric behavior.
We show that our approach can optimize the Laplacian of a surface-only mesh with random distant connections generated via \textsc{TetGen} \cite{tetGen2015} to approximate the spectral behavior of a volumetric mesh.

\begin{wrapfigure}[7]{r}{1.33in}
	\raggedleft
	\hspace*{-0.7\columnsep}
    \includegraphics[width=1.31in, trim={6mm 0mm 1mm 0mm}]{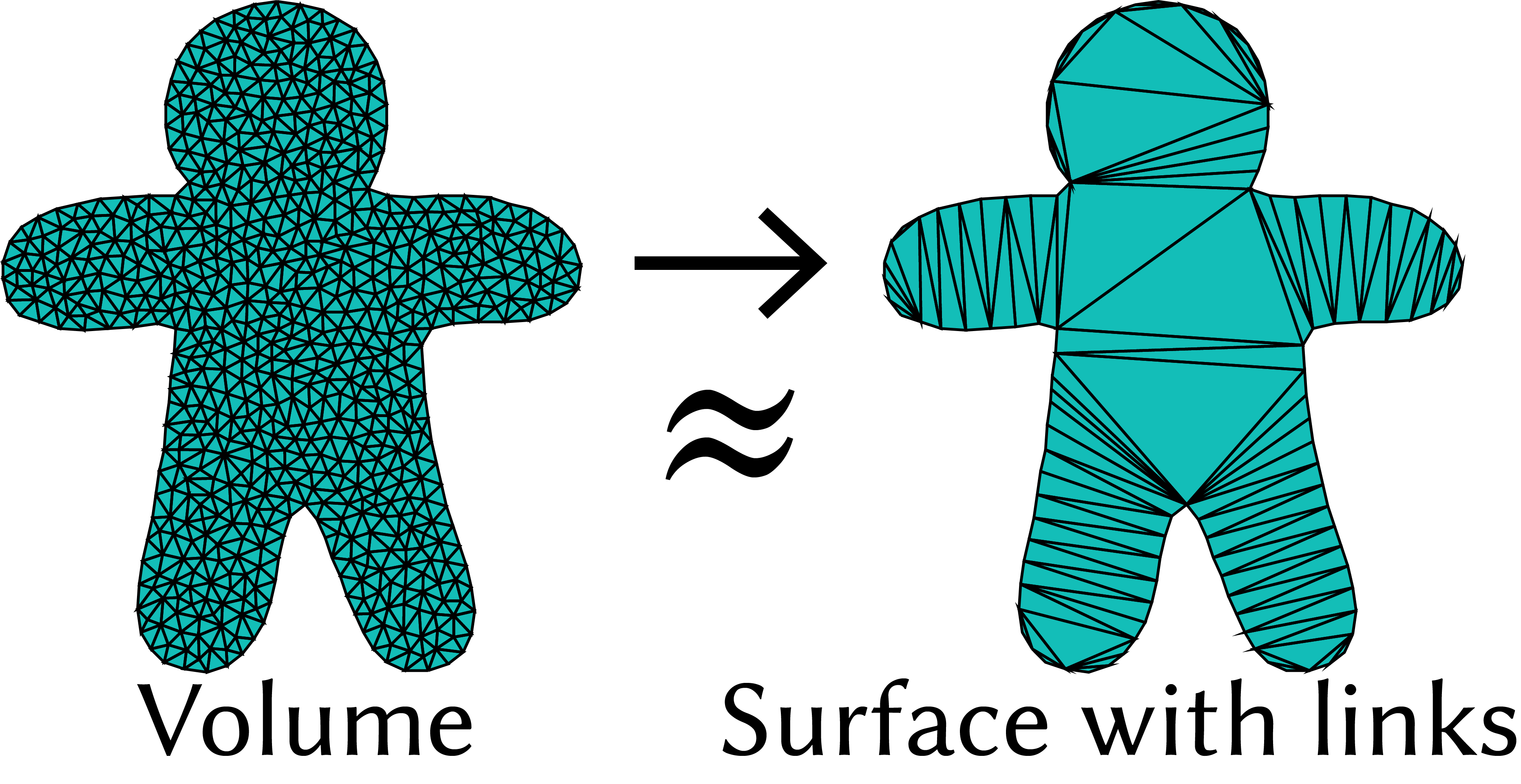}
	\label{fig:vol2surf_woody}
\end{wrapfigure} 
T\revise{aking} the boundary surface mesh of a volumetric tetrahedral mesh as the input, we first add distant edges to the surface Laplacian to determine the sparsity pattern.
%
%
%
We use the constrained Delaunay tetrahedralization in \textsc{TetGen} \cite{tetGen2015} to add the edges between ``visible'' but distant vertices (see inset), and use its pattern as the sparsity pattern of our modified surface Laplacian.
Then we optimize the modified operator to preserve the spectral behavior of the volumetric Laplacian.
Compared to the traditional discretization in Boundary Element Method \cite{ArtDefo1999} where the boundary matrices are usually dense, in our method the surface-only Laplacian can still remain sparse and maintain a similar sparsity pattern as its surface cotangent Laplacian.

In \reffig{vol2surf_craneHook} and \reffig{vol2surf_max_planck}, we visualize the functional map and biharmonic distance of our optimized surface Laplacian. We show that we can further capture the volumetric behavior by increasing the rings of neighborhood.

Similar to \cite{partialFMap2017}, our volume-to-surface mapping is also a partial functional mapping, which may lose some (internal) eigenvectors and result in a skewed functional map when the internal volume is large (see \reffig{vol2surf_max_planck}). 
Our method can also serve as a possible way to generate training data to find the best sparsity pattern without the presence of a volumetric mesh.
%

\subsection{Operator Detachment}
\begin{figure}
    \centering
    \includegraphics[width=3.33in]{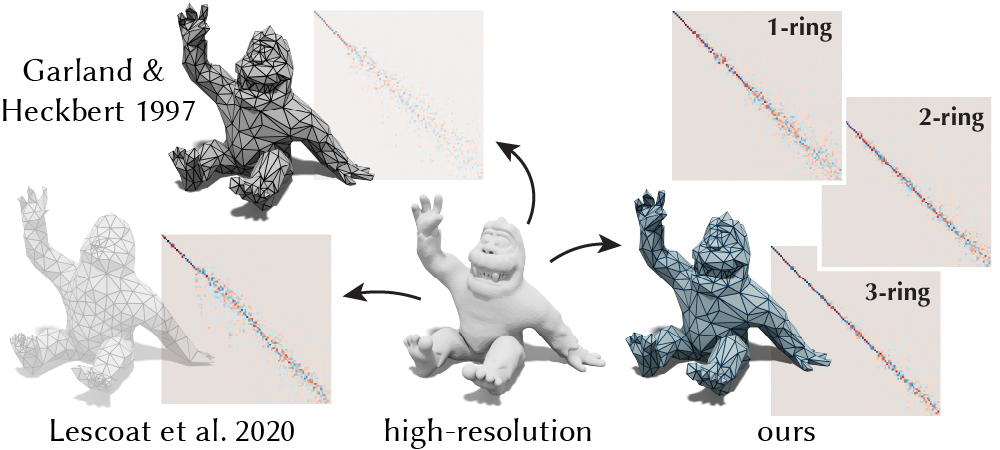}
    \caption{When one ties the differential operator with the mesh, we can either preserves the appearance of the mesh \cite{GarlandH97} or the spectral properties of the operator \cite{lescoat2020spectral}, but not both. Our approach enables one to detach the operator from the mesh (right) to achieve both simultaneously: using an appearance-preserving mesh for visualization and a spectral-preserving operator with user-desired sparsity patterns (\eg, 1-ring, 2-ring, or 3-ring) for computation.}
    \label{fig:detachOperator} 
    \vspace{-8pt}
\end{figure} 
\citet{sharp2019navigating} propose to represent the same geometry using two discrete representations: one for visualization and one for computation. In a similar spirit to \cite{sharp2019navigating}, our approach enables one to have one mesh for visualization and one detached operator for computation. 

Previous decimation methods either preserve the appearance but fail in preserving spectral properties or preserve the spectral properties but fail in preserving the appearance. 
This is partly due to the perspective of defining the operator directly on the discrete mesh, and partly due to the lack of tools to optimize the operator independently.  

In order to simultaneously preserve the appearance and the spectral properties, in \reffig{detachOperator} we first obtain a coarsened mesh from an appearance-preserving decimation, then we optimize the operator separately using the sparsity pattern defined by the connectivity of the mesh. 
Intuitively, this optimization tries to retrieve the desired properties on the original mesh by manipulating the metric ``seen'' by the coarsened operator. 
At the end of this process, even though the ``distorted'' metric may not be embeddable, one can always use the embeddable appearance-preserving mesh to visualize the results of the computation.
In \reffig{detachOperator}, this detachment allows us to preserve both the appearance and the spectral properties, while the method of \cite{GarlandH97} fails in preserving spectral properties and the method of \cite{lescoat2020spectral} fails in preserving the appearance.
In \reffig{teaser}, we demonstrate the strength of this approach in approximating the vibration modes of a high-resolution mesh using a coarse mesh with a detached coarsened operator. 
\revise{
Compared to \cite{liu2019spectral} which does not allow inputting an arbitrary sparsity pattern (instead it builds the output sparsity pattern by ``squaring'' an incidence matrix, see their Eq. 7), our method can take any sparsity pattern as input. 
This means one can geometrically simplify a mesh, then use that new mesh's sparsity pattern as input to our algorithm to optimize a compatible operator (see \reffig{teaser}), which enables its use in applications that require an embedded mesh and an accurate coarse operator (\eg, simulation with contact handling). }
\vspace{-2pt}
\section{Limitations \& Future Work}
\begin{figure}
  \centering
  \includegraphics[width=3.33in]{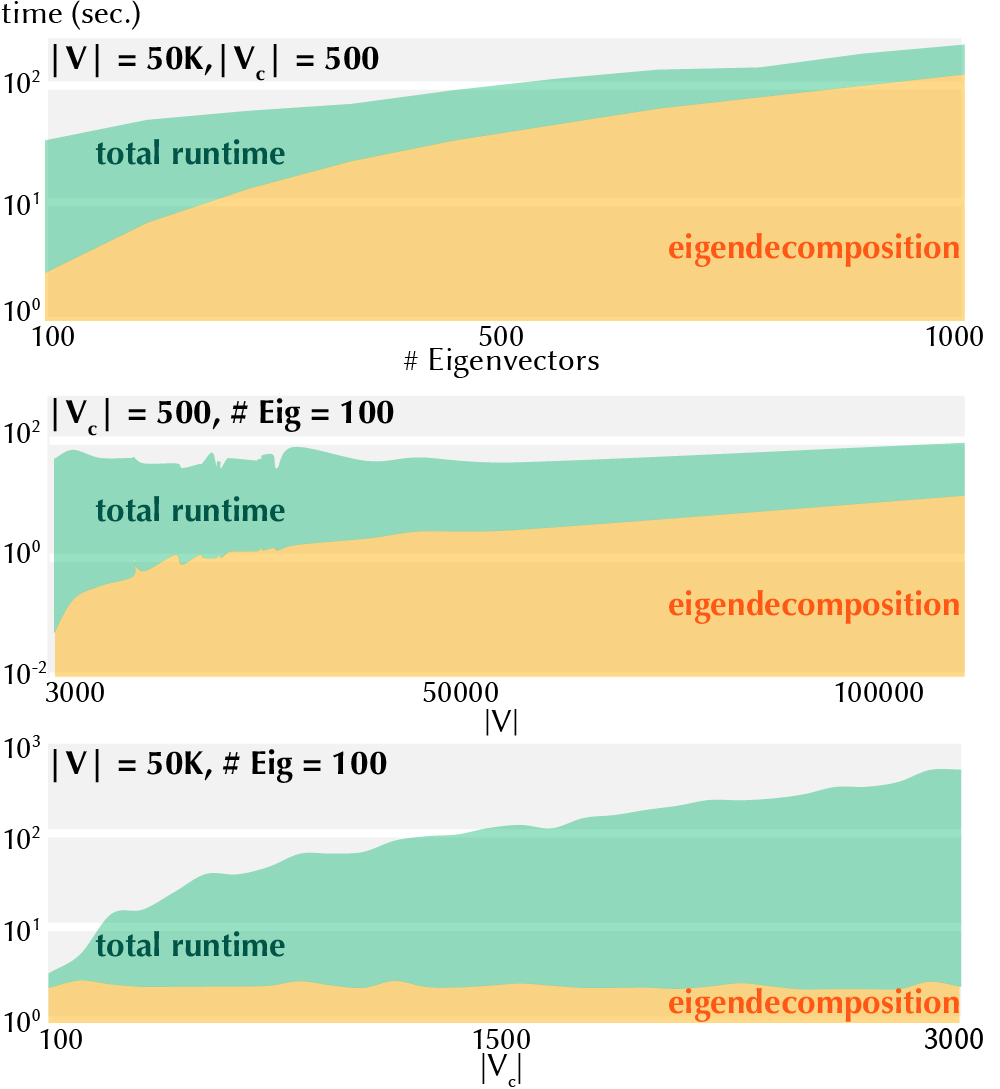}
  \caption{\revise{Our runtime shows that our method is more suitable for aggressive coarsening (middle). 
  When many eigenvectors are in use (top) or input meshes are large (middle), computing eigendecomposition can be the bottleneck.}}
  \label{fig:spec_coarse_runtime_eigen}
  \vspace{-12pt}
\end{figure} 
Further exploiting the limited degrees of freedom would enable an even better spectral preservation for 1-ring isotropic operator.
\revise{Jointly optimizing the sparsity pattern and the operator entries may lead to even finer solutions, especially for volume-to-surface approximation. }
%
%
Exploring different regularizers and energy formulations would be desirable for solving the underdetermined system when degrees of freedom are too large compared to the number of eigenvectors in use.
Avoid introducing additional low frequency eigenvectors 
\begin{wrapfigure}[7]{r}{0.4\linewidth}
  \vspace*{-0.5\intextsep}
  \hspace*{-0.5\columnsep}
  \begin{minipage}[b]{\linewidth}
  \includegraphics[width=\linewidth, trim={3mm 20mm 0mm 20mm}]{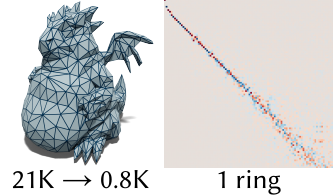}
  \label{fig:limitation_anisotropic}
  \end{minipage}
\end{wrapfigure}
during the optimization would benefit the downstream applications \revise{(see the inset)}.
\revise{Reducing the memory consumption of the Kronecker product would further increase the scalability of our method (see \refapp{kronecker_product}). }
%
%
Incorporating a fast eigen-approximation or removing the use of eigen decomposition would further accelerate the spectral coarsening \revise{(see \reffig{spec_coarse_runtime_eigen})}.
Further analysis of the tradeoff between \revise{the convergence and the number of cliques} could offer insight towards future applications of chordal decomposition.
%
Extending our spectral coarsening of surface-based geometric operators to volumetric stiffness matrix could also provide an alternative way to deal with the numerical stiffening in simulation.
\revise{As} a first order method, ADMM is slow to obtain highly accurate solutions, but fast in getting moderately accurate solutions.
Similar to other splitting methods, ADMM is sensitive to the conditioning of the problem data. 
Thus adding a preconditioner could make our solver more robust to the scaling problem and increase its performance. 
\revise{
Finally, it would be also interesting to extend our method to many other applications beyond geometry processing and shape matching, such as physics-based simulation, topology optimization, algebraic multigrid and spectral graph reduction. }








\begin{acks}
This work is funded in part by NSERC Discovery (RGPIN-2017-05524, RGPIN-2017–05235, RGPAS–2017–507938), Connaught Fund (503114), CFI-JELF Fund, Accelerator (RGPAS-2017-507909), New Frontiers of Research Fund (NFRFE–201), the Ontario Early Research Award program, the Canada Research Chairs Program, the Fields Centre for Quantitative Analysis and Modelling and gifts by Adobe Systems, Autodesk and MESH Inc.
We especially thank Yifan Sun, Giovanni Fantuzzi and Yang Zheng for their enlightening discussions and advice about the chordal decomposition, and Thibault Lescoat for sharing the spectral simplification implementation and discussions about running experiments.
We thank Abhishek Madan, Silvia Sell\'an, Michael Xu, Sarah Kushner, Rinat Abdrashitov, Hengguang Zhou and Kaihua Tang for proofreading;
Mirela Ben-Chen for insightful discussions about the weighted functional map; 
Josh Holinaty for testing the code;  
John Hancock for the IT support; 
anonymous reviewers for their helpful comments and suggestions.
\end{acks}

\bibliographystyle{ACM-Reference-Format}
\bibliography{sections/reference.bib}

\appendix

\section{Alternating direction method of multipliers}\label{app:ADMM}
Alternating direction method of multipliers (ADMM) solves optimization problems in the following format
\begin{alignat}{2}
    &\min_{\vx,\vz}\  &&f(\vx) + g(\vz)\\
    & \text{ s.t.}\  && \mA\vx + \mB\vz = \vc.
\end{alignat}
The (scaled) ADMM solves the problem by iteratively applying the following steps
\begin{align}\label{equ:ADMMSteps}
    &\vx^{t+1}  \coloneqq \argmin_\vx \Big( f(\vx) + \frac{\rho}{2} \| \mA\vx + \mB\vz^t - \vc + \vu^t \|^2_2 \Big)  \nonumber \\
    &\vz^{t+1}  \coloneqq  \argmin_\vz \Big( g(\vz) + \frac{\rho}{2} \| \mA\vx^{t+1} + \mB\vz - \vc + \vu^t \|^2_2 \Big) \\
    &\tilde{\vu}^{t+1} \coloneqq \vu^t + \mA\vx^{t+1} + \mB\vz^{t+1} - \vc \nonumber \\
    &\rho^{t+1}, \vu^{t+1} \coloneqq \text{update}(\rho^t), \nonumber
\end{align}
where $\rho$ is the \emph{penalty parameter} and $\vu$ is \revise{the} \emph{scaled dual variable}. In the last step, a common strategy is to update the penalty $\rho$ as 
\begin{align}
	\rho^{t+1} = \begin{cases}
		\tau^\text{incr} \rho^t & \text{if $\| \vr^t\|_2 > \mu \| \vs^t\|_2$}\\
		\rho^t / \tau^\text{decr} & \text{if $\| \vs^t\|_2 > \mu \| \vr^t\|_2$}\\
		\rho^t & \text{otherwise},
	\end{cases}
\end{align}
where $\tau^\text{incr} > 1$, $\tau^\text{decr} > 1$, $\mu > 1$ are parameters\revise{,} $\vr$ and $\vs$ are the \emph{primal residual} and the \emph{dual residual}, respectively. We can compute them as
\begin{align}
	\vr^{t+1} = \mA\vx^{t+1} + \mB\vz^{t+1} - \vc, \quad \vs^{k+1} = \rho \mA^\top \mB (\vz^{t+1} - \vz^{t}).
\end{align}
After updating $\rho$ we must also scale the dual variable $\vu$ as
\begin{align}
	\vu^{t+1} = \tilde{\vu}^{t+1} \times \frac{\rho^t}{\rho^{t+1}}.
\end{align}
A common stopping criteria is when both $\| \vr^t\|_2 < \epsilon^\text{pri}$ and $\| \vs^t\|_2 < \epsilon^\text{dual}$ are below the thresholds $\epsilon^\text{pri}, \epsilon^\text{dual}$. We only review basic concepts of ADMM \revise{here} for \revise{self-containedness}. \revise{We} wholeheartedly refer the reader to a great survey \cite{boyd2011ADMM} for more information on ADMM.

\section{Change of Variables} \label{app:changeOfVariables}
We describe the details on how to apply change of variables and vectorization for the constraints presented in \refequ{opt2}.

Given the matrices $\P_\E$ and $\invP_\E$ in \refequ{x2xE}, which allow us to go back and forth between $\text{vec}(\X)$ and $\xE$, we can vectorize the equality constraint in \refequ{opt2} as
\begin{align}
    \text{vec}(\X \vv)= \text{vec}(\ve)\ 
    &\ \Rightarrow\ (\vv^\top \otimes \mI )\ \text{vec}(\X) = \ve \\
    &\ \Rightarrow\ \underbrace{(\vv^\top \otimes \mI ) \invP_\E}_{\mG}\ \xE = \ve \\
    &\ \Rightarrow\ \mG \xE = \ve,
\end{align}
where $\mI$ is the identity matrix. For the chordal decomposition \refequ{opt2_chordal}, we can directly apply the vectorization strategy discussed in \refsec{chordalDecomposition} as
\begin{align}
    \text{vec}(\X) = \sum_{\i=1}^p \text{vec}(\P_\i^\top \Z_\i \P_\i) 
    &\ \Rightarrow\  \text{vec}(\X) = \sum_{\i=1}^p \K_\i \underbrace{\text{vec}(\Z_\i)}_{\z_\i}\\
    &\ \Rightarrow\ \invP_\E \xE = \sum_{\i=1}^p \K_\i \z_\i,
\end{align}
where we define $\z_\i \coloneqq \text{vec}(\Z_\i)$. Therefore we can easily rewrite the PSD constraint on $\Z_\i$ as
\begin{align}
    \Z_\i = \text{vec}^{\text{-}1}(\z_\i) \in \S_+^{n_\i}.
\end{align}
Combining all these results gives us the formulae in \refequ{opt3}.

\section{Derivation of ADMM Steps}\label{app:admmForm}
Here we describe how to derive the ADMM steps (see \refequ{ADMMSteps}) to solve the optimization in \refequ{reduce_opt3}. Our derivation follows a similar strategy described in Sec. 4.2 \cite{Zhen2020ADMMSDP}.

We start by introducing an auxiliary variable $\y$ such that 
\begin{alignat}{3}
    &\min_{\xE,\y,\rz}\  &&f(\xE) \label{equ:opt4} \\
    & \text{ s.t.}\  &&\mG \xE = \ve \\
    & && \P_\C \invP_\E \xE = \P_\C \rK \y \\
    & && \text{vec}^{\text{-}1}( \Q_\i \rz_\i)  \succeq 0, \quad &&\i = 1,\cdots, p  \\
    & && \y = \rz, 
\end{alignat}
Then we introduce the indicator function $\delta_\mathcal{W}$ as 
\begin{align}
    \delta_\mathcal{W}(x) = \begin{cases}
        0, & x \in \mathcal{W} \\
        \infty, & \text{otherwise}
    \end{cases}.
\end{align}
This allows us to \revise{rewrite} \refequ{opt4} as
\begin{alignat}{3}\label{equ:opt5}
    &\min_{\xE,\y,\z}\  &&\underbrace{f(\xE) + \delta_\ve(\mG \xE) + \delta_\textbf{0}(\P_\C \invP_\E \xE - \P_\C \rK \y)}_\text{function of $\mathcal{X} = \{ \xE,\y \}$} \nonumber\\
    & && \qquad + \underbrace{\sum_{\i=1}^p \delta_{+} \big(\text{vec}^{\text{-}1}(\Q_\i \rz_\i)\big)}_\text{function of $\mathcal{Z} = \{ \rz \}$} \nonumber\\
    & \text{ s.t.}\  && \y - \rz = 0,
\end{alignat}
where we use $\delta_{+}$ to denote the indicator function for the PSD constraint.
This \revise{format} of the optimization enables us directly apply the ADMM step \refequ{ADMMSteps}. In particular the update of $\mathcal{X} = \{ \xE,\y \}$ is as follows 
\begin{alignat}{2}\label{equ:argminX}
    &\argmin_{\xE,\y}\  && f(\xE) + \frac{\rho}{2} \| \y - \rz^t + \vu^t \|^2_2 \nonumber\\
    & \text{ s.t.}\  && \mG \xE = \ve  \\ 
    & && \P_\C \invP_\E \xE - \P_\C \rK \y = 0,\nonumber
\end{alignat}
where the solution depends on the energy \revise{function} $f$ in use. In the case of spectral coarsening energy, this boils down to a single linear solve of the KKT system (see \refapp{argminX_spectralCoarsening}). The update of $\mathcal{Z} = \{ \rz \}$ is 
\begin{alignat}{2} 
    &\argmin_{\rz}\  && \sum_{\i=1}^p \| \y^{t+1}_\i - \rz_\i + \vu_\i^t \|^2_2\\
    & \text{ s.t.}\  && \Q_\i \rz_\i \succeq 0 \quad \i = 1,\cdots, p.
\end{alignat}
This can be solved by projecting a set of small dense matrices $\text{vec}^{\text{-}1}\big(\Q_\i (\y^{t+1}_\i + \vu^t_\i )\big)$ to PSD, which requires \revise{us} to solve the eigen-decomposition and remove the negative eigenvalues. Note that this process can be solved efficiently because each matrix to be projected is small and this process can be trivially parallelized. 


\section{\texorpdfstring{$\argmin_{\mathcal{X}}$}{argminX} for Spectral Coarsening}\label{app:argminX_spectralCoarsening}
Applying ADMM to solve \revise{the} spectral coarsening \revise{problem} requires \revise{us} to derive the update on $\mathcal{X}$ (see \refequ{argminX}). We start by vectorizing the spectral coarsening energy \refequ{spectralCoarseningEnergy} as
\begin{align}
    f(\X) &= \frac{1}{2} \|\mR \mM^{\text{-}1}\mL \Phi - \widetilde{\mM}^{\text{-}1} \X \mR \Phi \|^2_{\widetilde{\mM}} \\
    &= \frac{1}{2} \|\underbrace{\widetilde{\mM}^{\nicefrac{1}{2}} \mR \mM^{\text{-}1} \mL \Phi}_{\mW} - \underbrace{\widetilde{\mM}^{\nicefrac{-1}{2}}}_{\mV}\ \X\ \underbrace{\mR \Phi}_{\mU} \|^2_F \\
    &= \frac{1}{2} \| \mW - \mV \X \mU \|^2_F \\
    &= \frac{1}{2} \| \text{vec}(\mW) -  \text{vec}(\mV \X \mU) \|^2_2 \\
    &= \frac{1}{2} \| \text{vec}(\mW)-  (\mU^\top \otimes \mV)\ \text{vec}(\X) \|^2_2.
\end{align}
We then apply change of variables in \refequ{x2xE} to modify the energy as follows
\begin{align} \label{equ:vecSpectralEnergy}
    f(\xE) &= \frac{1}{2} \| \underbrace{\text{vec}(\mW)}_{\vw}-  \underbrace{(\mU^\top \otimes \mV) \invP_\E}_{\mE} \xE \|^2_2 
    =  \frac{1}{2} \|  \vw - \mE \xE \|^2_2.
\end{align}

Updating $\mathcal{X} = \{ \xE, \y \}$ in the ADMM (\refequ{argminX}) amounts to obtaining the minimizer of the following problem
\begin{alignat}{2}
    &\min_{\xE,\y}\  && \frac{1}{2} \|  \vw - \mE \xE \|^2_2 + \frac{\rho}{2} \| \y - \rz^t + \vu^t \|^2_2 \\
    & \text{ s.t.}\  && \mG \xE = \ve,  \\ 
    & && \underbrace{\P_\C \invP_\E}_\mC \xE - \underbrace{\P_\C \rK}_\mD \y = 0.
\end{alignat}
We first derive the Lagrangian with multipliers $\mu_1, \mu_2$ as
\begin{align}
    \mathcal{L}(\xE, \y, \mu_1, \mu_2) &= \frac{1}{2} \|  \vw - \mE \xE \|^2_2 + \frac{\rho}{2} \| \y - \rz^t + \vu^t \|^2_2 \\
    &\qquad + \mu_1^\top(\mC \xE - \mD \y) + \mu_2^\top(\mG \xE).
\end{align}
Setting the derivatives to zeros gives us
\begin{align}
    \frac{\partial \mathcal{L}}{\partial \xE} = 0 \ &\Rightarrow \ \mE^\top \mE \xE + \mC^\top \mu_1 + \mG^\top \mu_2  = \mE^\top \vw\\
    \frac{\partial \mathcal{L}}{\partial \y} = 0\ &\Rightarrow \ \y = \rz - \vu + \frac{1}{\rho} \mD^\top \mu_1,\\
    \frac{\partial \mathcal{L}}{\partial \mu_1} = 0\ &\Rightarrow \ \mC \xE - \mD \y = 0,\\
    \frac{\partial \mathcal{L}}{\partial \mu_2} = 0\ &\Rightarrow \ \mG \xE = 0.
\end{align}
We can substitute the expression of $\y$ from $\nicefrac{\partial \mathcal{L}}{\partial \y} = 0$ to $\nicefrac{\partial \mathcal{L}}{\partial \mu_1} = 0$ and then obtain a set of equations
\begin{align}
    &\mE^\top \mE \xE + \mC^\top \mu_1 + \mG^\top \mu_2  = \mE^\top \vw\\
    &\mC \xE - \frac{1}{\rho} \mD \mD^\top \mu_1 = \mD (\rz - \vu)\\
    &\mG \xE = 0.
\end{align}
This enables us to obtain the optimal $\xE^\star, \mu_1^\star$ via solving a linear system
\begin{align}\label{equ:KKT_Thm2}
    \begin{bmatrix}
        \mE^\top \mE & \mC^\top & \mG^\top \\
        \mC & \nicefrac{-1}{\rho}\ \mD \mD^\top & \textbf{0} \\
        \mG & \textbf{0} & \textbf{0}
    \end{bmatrix} 
    \begin{bmatrix}
       \xE \\
        \mu_1 \\
        \mu_2
    \end{bmatrix} 
    =
    \begin{bmatrix}
        \mE^\top \vw  \\
        \mD (\rz - \vu) \\
        \textbf{0}
     \end{bmatrix}.
\end{align}
Then we can recover the optimal $\y^\star$ as
\begin{align}
    \y = \rz - \vu + \frac{1}{\rho} \mD^\top \mu_1^\star.
\end{align}

\section{\texorpdfstring{$\argmin_{\mathcal{X}}$}{argminX} for Spectral Coarsening}\label{app:argminX_spectralCoarsening}
When the number of eigenvectors that are chosen to preserve is large, the size of $\mU^\top \otimes \mV$ in \refequ{vecSpectralEnergy} can be large.
However, we can avoid explicitly construct $\mU^\top \otimes \mV$ by leveraging the fact that only $\mE^\top \mE$ and $\mE^\top \vw$ are used in the linear solve \refequ{KKT_Thm2}.
By using the properties $(\mA \otimes \mB)(\mC \otimes \mD) = (\mA\mC) \otimes (\mB\mD)$ and $(\mB \otimes \mA)\text{vec}(\mX) = \text{vec}(\mA\mX\mB)$, we can instead compute $\mE^\top \mE$ and $\mE^\top \vw$ as
\begin{align}
    \mE^\top \mE & = ((\mU^\top \otimes \mV) \invP_\E)^\top (\mU^\top \otimes \mV) \invP_\E \\
     & = (\invP_\E)^\top (\mU^\top \otimes \mV)^\top (\mU^\top \otimes \mV) \invP_\E \\
     & = (\invP_\E)^\top (\mU \otimes \mV^\top) (\mU^\top \otimes \mV) \invP_\E \\
     & = (\invP_\E)^\top ((\mU \mU^\top) \otimes (\mV^\top \mV)) \invP_\E, \\
    \mE^\top \vw & = ((\mU^\top \otimes \mV) \invP_\E)^\top \text{vec}(\mW) \\
     & = (\invP_\E)^\top (\mU \otimes \mV^\top) \text{vec}(\mW) \\
     & = (\invP_\E)^\top \text{vec}(\mV^\top \mW \mU^\top),
\end{align}
where the size of $(\mU \mU^\top) \otimes (\mV^\top \mV)$ and $\mV^\top \mW \mU^\top$ are independent of the number of eigenvectors we choose to preserve.



\section{Implementation} \label{app:implementation}
Our solver is implemented in MATLAB using gptoolbox \cite{gptoolbox}. 
We adapt the MATLAB code from \cite{sun2015decomposition} to compute the chordal decompoistion. 
Runtimes for all the examples were reported on a MacBook Pro with an Intel i5 2.3GHz processor, 16GB of RAM and an Intel Iris Plus Graphics 655 GPU. 
Experiments for volume to surface were tested on a Linux workstation with an Dual 14 Core 2.2Ghz processor, 383GB of RAM and 2 Titan RTX 24GB GPU.
We did not use multi-threading, though the projection to PSD cones can be easily parallelized using MATLAB MEX file.
Since the KKT system matrix in $\argmin_\X$ remains the same until $\rho$ is updated, we only perform numerical factorization when $\rho$ is updated and reuse it until $\rho$ changes again (usually after tens of iterations). 

For consistency, we choose to evaluate all the results on the first 100 eigenvectors across the experiments unless specified otherwise.
We preserve the first 100 eigenvectors for surface Laplacian in spectral coarsening and simplification, and use an increased number of eigenvectors for volumetric Laplacian or when the system goes underdetermined.
We also normalize all the eigenvectors to have unit length and scale the mesh to ensure each vertex \revise{has} unit area.
For a fair comparison, we compare the runtime of our MATLAB implementation with the MATLAB implementation of \cite{liu2019spectral}.
When comparing against \cite{lescoat2020spectral}, we use their decimation algorithm without edge flips and enable approximation of the minimizer on collapse edges.

\begin{figure}
    \centering
    \includegraphics[width=3.33in]{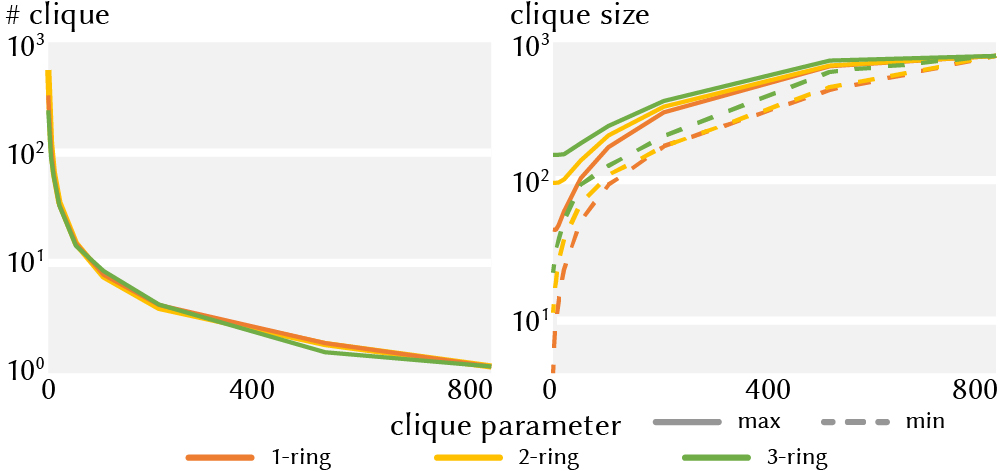}
    \caption{\revise{We plot the change of the average number of cliques and the average maximal and minimal clique size with respect to clique parameters which we control in the chordal decomposition algorithm when coarsening various meshes to 800 vertices. }}
    \label{fig:clique_num_size_123ring}
    \vspace{-12pt}
\end{figure} 
\begin{figure}
    \centering
    \includegraphics[width=3.33in]{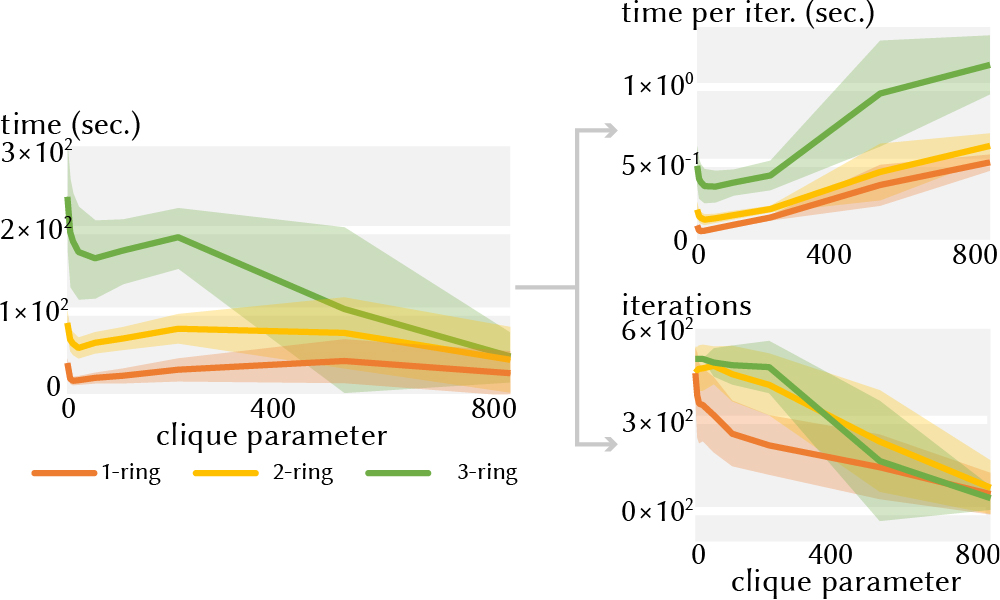}
    \caption{\revise{We show the change of the total ADMM runtime, the ADMM runtime per iteration and the number of iterations with respect to clique parameters when coarsening a number of meshes to 800 vertices. Here the lines denote the average and the color regions denote the standard deviation. 
    %
    }}
    \label{fig:clique_time_123ring}
    \vspace{-12pt}
\end{figure} 
In our implementation, the projecting of each clique matrix to PSD is relatively cheap because the size of clique matrix usually varies from tens to a few hundreds and can be controlled by the parameters during the clique merging stage of chordal decomposition.
The size of the clique matrix after the chordal decomposition would be approximately around the clique merging parameters.
In our \revise{experiments}, we set the parameters for clique merging to be 200 so that the size of the clique matrix is in a few hundreds \revise{considering} the tradeoff between eigendecomposition speed and convergence rate.
\revise{As shown in \reffig{clique_num_size_123ring} and \reffig{clique_time_123ring}, there is a non-monotonic relationship between the clique parameter and the ADMM runtime, and we experimentally determine that a parameter of 200 works best for all our examples. Optimal parameter determination is left for future work. }
We recommend setting \revise{the clique parameters} to be larger than 100 when only preserving the first 100 eigenvectors \revise{to ensure our method converges}.
We also notice that increasing the number of eigenvectors preserved would lead to better convergence and avoid underdeterminism in the system. 
Let $k$ be the number of the eigenvectors we choose to preserve and $m$ be the number of vertices in the coarsened domain.
When the DOF defined by the sparsity pattern is large (\ie, volumetric Laplacian, 3-ring surface Laplacian) , we recommend setting the number of the preserved eigenvectors to be $k > 0.5 * m$, and \revise{using} the weighted energy to preserve the low-frequency modes.
Experimentally, we observe that when the DOF is too large, the system may become underdetermined for volumetric mesh and 2- or 3-ring if $m>2 \times k$.

\section{Additional Results}
\begin{figure}
    \centering
    \includegraphics[width=3.33in]{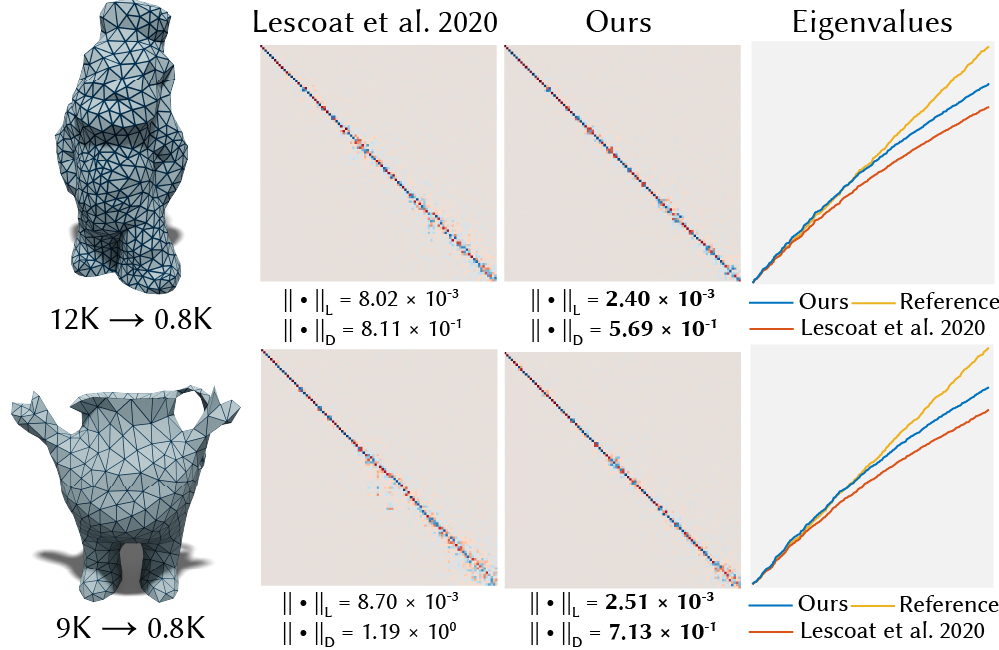}
    \caption{Our algorithm can further improve the spectral properties of \cite{lescoat2020spectral} as a post-processing step.
    }
    \label{fig:spec_simpl_refine_chess_cupman}
    \vspace{-8pt}
\end{figure} 
\begin{figure}
    \centering
    \includegraphics[width=3.33in]{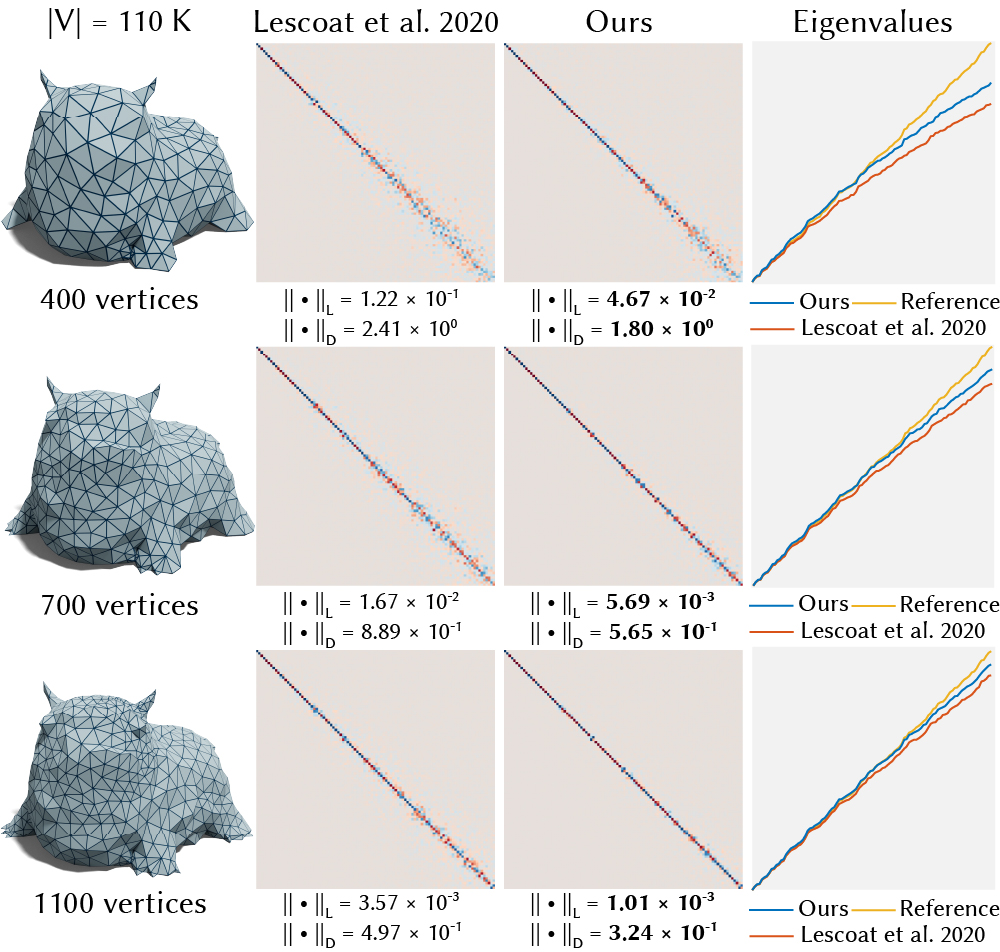}
    \caption{Due to the freedom of choosing the output sparsity pattern, our method can serve as a post-processing tool to further improvze the resulting operator from the method of \cite{lescoat2020spectral}. The results indicate that our post-processed operators result in better functional maps (middle) compared to the output operators from \cite{lescoat2020spectral} (left), so as the eigenvalues (right). }
    \label{fig:spec_simpl_compare_monster_110K}
    \vspace{-10pt}
\end{figure}
\begin{figure}
    \centering
    \includegraphics[width=3.33in]{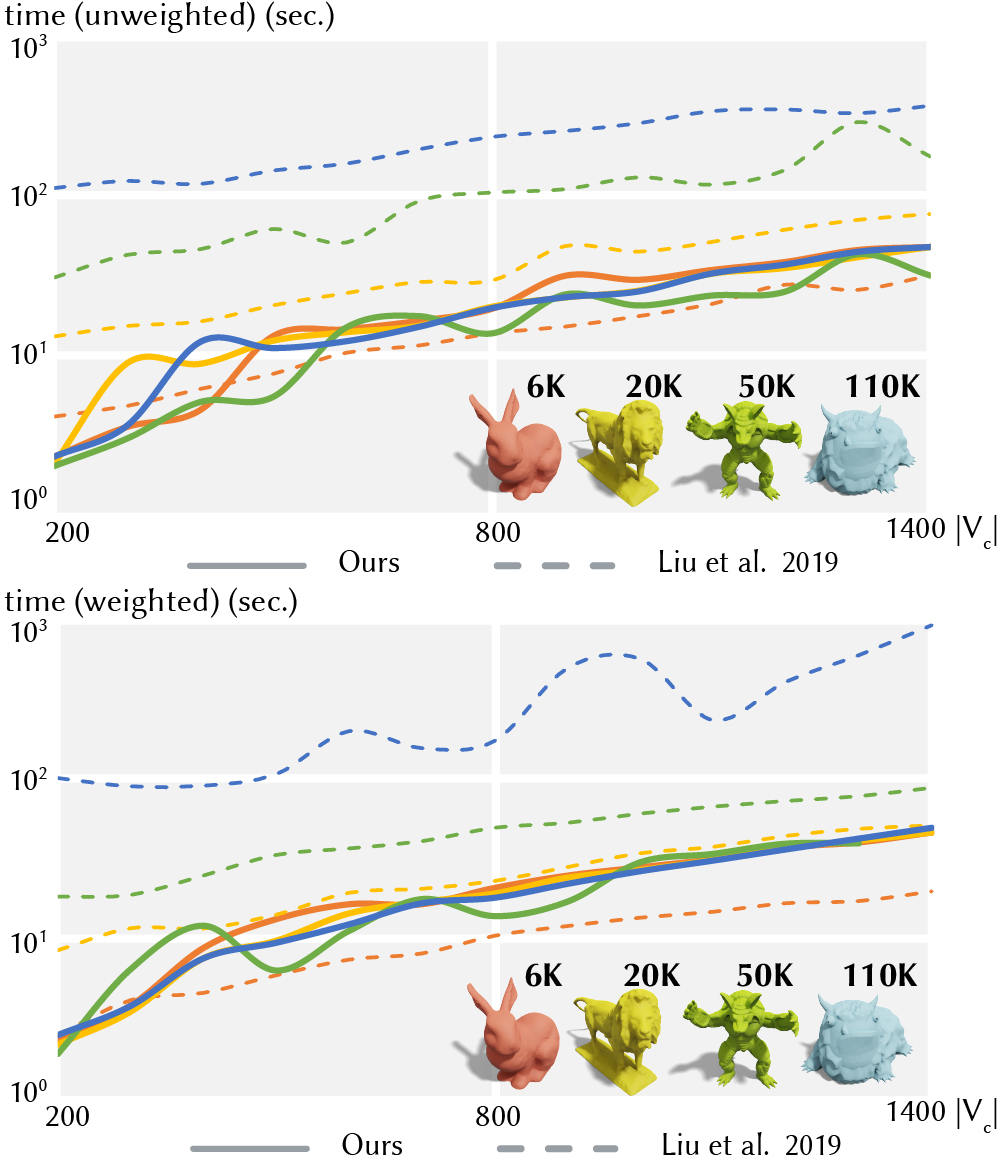}
    \caption{We compare the runtime of our optimization algorithm of both the weighted and unweighted version with \cite{liu2019spectral} using the same 3-ring sparsity pattern.
    Here we only consider the solve time, factoring out the precomputation for both our method and the method of \cite{liu2019spectral}.
    }
    \label{fig:spec_coarse_compare_time_armadillo_both}
    \vspace{-8pt}
\end{figure}

 
In addition to the results in \refsec{results}, we report more results on the spectral simplification in \revise{\reffig{spec_simpl_refine_chess_cupman} and \reffig{spec_simpl_compare_monster_110K}} and an extended evaluation on runtime in \reffig{spec_coarse_compare_time_armadillo_both} to complement the main text. 

\end{document}